\documentclass[11pt]{article}
\usepackage[a4paper,margin=25mm]{geometry}
\usepackage{graphicx}
\usepackage{amsmath,amssymb}
\usepackage{mathtools}
\usepackage{bm}
\usepackage[numbers,sort&compress,super]{natbib}
\usepackage{xcolor}
\usepackage{caption}
\usepackage{microtype}
\usepackage[colorlinks=true, allcolors=blue]{hyperref}
\usepackage{authblk}
\newcommand{\md}{\mathrm{d}}
\def\kb{k_{\text B}}
\def\kon{k_{\text{on}}}
\def\koff{k_{\text{off}}}

\title{A Transient-Network Model of Condensate Aging and Morphology Transition in Phase Separation}

\author[1]{Bhanjan Debnath}
\affil[1]{Department of Chemical Engineering, Indian Institute of Technology Hyderabad, Kandi, Sangareddy, Telangana 502285, India}
\affil[ ]{\texttt{bhanjan@che.iith.ac.in}}
\date{}

\newcommand{\significancetext}{Biomolecular condensates play important roles in intra- and inter-cellular organization and signaling through maintenance of their material properties and spatial morphology. Liquid-like condensates can coarsen, whereas aging condensates resist coarsening because elastic memory progressively builds up, leading to arrested structures. This work presents a phase-separation model coupled with transient-network theory. A network is formed by protein molecules switched to a conformational state that exposes additional cross-linking domains. Cross-link dynamics dictate network fluidity, and the cross-links can stabilize as aging progresses. The model explains how molecular switching and cross-link survival regulate the mechanics of network-like phases and translate into morphological transitions during phase separation.}

\newcommand{\keywordtext}{phase separation; transient network; biomolecular condensates; dynamic cross-linking; aging}

\begin{document}
\maketitle

\begin{abstract}
Biomolecular condensates can undergo striking changes, such as transitioning from a liquid-like to a gel- or a solid-like aggregate due to changes in molecular interactions in response to changes in the biochemical environment. The question of how modified molecular interactions lead to such a transition in the material properties and spatial organization of condensates has not yet been elucidated. To address this question, we represent the biochemical environment as a triphasic mixture comprising a liquid-like protein-rich phase, a network-like protein-rich phase, and solvent. Owing to a change in the biochemical environment, protein molecules can reversibly switch between two conformational states. In a switched conformational state, the cross-linking domains of molecules are exposed which promote transient network formation in phase separated states. We develop a transient-network model and a continuum framework that couples phase separation, molecular switching, and dynamic cross-linking to predict condensate morphology and mechanics. The transient-network model predicts that a non-aging network behaves like a Maxwell fluid. When a network slowly ages via stabilization of cross-links, it shows  Maxwell-like behavior and waiting time-dependent relaxation. However, a strongly aged network shows elastic recoil like characteristic of a  Kelvin-Voigt solid. Our coupled continuum model demonstrates that the interplay of molecular switching and 
dynamic cross-linking in network formation shapes the spatial organization of condensate phases. In summary, this work demonstrates a mechanistic route explaining how conformational switching and molecular cross-linking  regulate material properties and morphology of condensates.
\end{abstract}


\section*{Significance}
\significancetext

\section{Introduction}
The active biochemical environment inside cells encourages the compartmentalization of proteins and biomolecules, and also continually remodels their material properties \citep{lyon2021framework,alberti2019considerations}. Maintaining material properties of condensates inside cells is essential for multiple purposes, such as signaling \citep{arimoto2008formation,costa2020integrated}, membrane structural adaptability \citep{bussi2023stress,wang2024biomolecular,sanfeliu2025mechanobiology}, maintaining fluidity of the cytoplasm \citep{xie2024polysome} and controlling electrochemical functions \citep{dai2024unlocking}, etc. Similar to demixing in a liquid solution, liquid-liquid phase separation (LLPS) can explain condensate formation \citep{hyman2014liquid,choi2020physical}, but does not necessarily determine the fate of condensates. It is well established that phase-separated condensates do not always remain in a liquid-like state. In many condensate systems, earlier work reported a gradual decrease in the fluidic nature of liquid-like condensates as they mature and age \cite{patel2015liquid, jawerth2020protein,linsenmeier2022dynamic,Alshareedah2024}. Progressive reduction in the fluidic nature drives condensates towards gel- or solid-like aggregates by reducing molecular mobility. Such transitions in which condensates lose their physiological function lead to cellular dysfunction and diseases \citep{alberti2021biomolecular, pei2025transcription}. 

It has been observed that factors such as perturbations in the  biochemical environment (for example, change in pH \citep{jin2022effects, yu2025aging}, depletion of ATP \citep{saurabh2022atp,linsenmeier2022dynamic} and loss of chaperone activity \citep{Yoo2022Chaperones,bard2024chaperone}), presence of prions \citep{gilks2004stress}, mutation \citep{patel2015liquid} and external stress \citep{arimoto2008formation,saha2025role}  change molecular interactions and can accelerate irreversible transition of condensates from liquid-like to solid-like states. Such transitions severely affect the morphology of condensates by arresting early stage coarsening and late-stage  dissolution upon  restoration of homeostatic cellular conditions. It is indicative that both the morphology and material properties that dictate the mechanical response of condensates in a perturbative environment can play a role in biomarkers. Therefore, it is of crucial importance to explore how altered molecular interactions lead to transitions in mechanical properties and modify morphology of condensates.

A newly formed condensate may initially be in a liquid-like phase, allowing neighbor exchanges, coarsening and dissolution \citep{zhang2024exchange,saha2025role}. However, if lifetimes of molecular interactions increase, elastic memory can progressively build up in the same condensate, resulting in suppression in  coarsening and neighbor exchanges and resistance to dissolution in the late stage. 
Until now, several attempts have been made to explain either how condensates form and coarsen \citep{tanaka1995new,tanaka2000viscoelastic,shimizu2015novel, adame2020liquid, kirschbaum2021controlling,shrinivas2021phase,erkamp2023spatially, julicher2024droplet,rossetto2025binding,paulin2026dynamics} or how elastic memory can build up while aging \citep{jawerth2020protein,lin2022modeling, takaki2023theory,blazquez2023location,Alshareedah2024,biswas2024molecular,espejo2025compositional}. However, an important question remains: what mechanism can drive phase separation and the subsequent development of elastic memory in phase-separated condensates, leading to long time-consuming coarsening and dynamic arrest of morphology?  
Against this backdrop, we  develop a continuum model that serves to highlight  morphological and mechanical transitions resulting from alteration in molecular interactions. 

A body of work encourages us to hypothesize that protein molecules having intrinsically disordered regions (IDRs) can switch to a different conformational state due to the presence or absence  of other biomolecules or ions \citep{guillen2020rna,morishita2023sodium,leder2025multichaperone}. In the switched state, IDRs of proteins are more exposed, can act as cross-links and favorably form  a transient network-like phase when they phase separate \citep{dar2024biomolecular}. If these cross-links stabilize over time through increasing structural ordered-ness or, {\it vice-versa}, fluidity and transient nature of such networks reduce as time passes, indicating aging. To predict such features, we develop a transient network model and determine the  factors that affect the network mechanics. Our model captures the experimentally observed transitions from Maxwell stress relaxation to Kelvin-Voigt elastic recoil when a network-like condensate phase progressively ages. We then couple the transient network model with a continuum model to examine how network-like phase formation impacts the coarsening dynamics and morphology of both liquid-like and network-like phases. In doing so, our model demonstrates a plausible mechanistic bridge between molecular switching and the emergent features of condensate dynamics. 

\section{Model}
\label{sec:model1}

\subsection{Free energy formulation}
\label{sec:mod_1}

\begin{figure}
\centering
\includegraphics[width=0.7\linewidth]{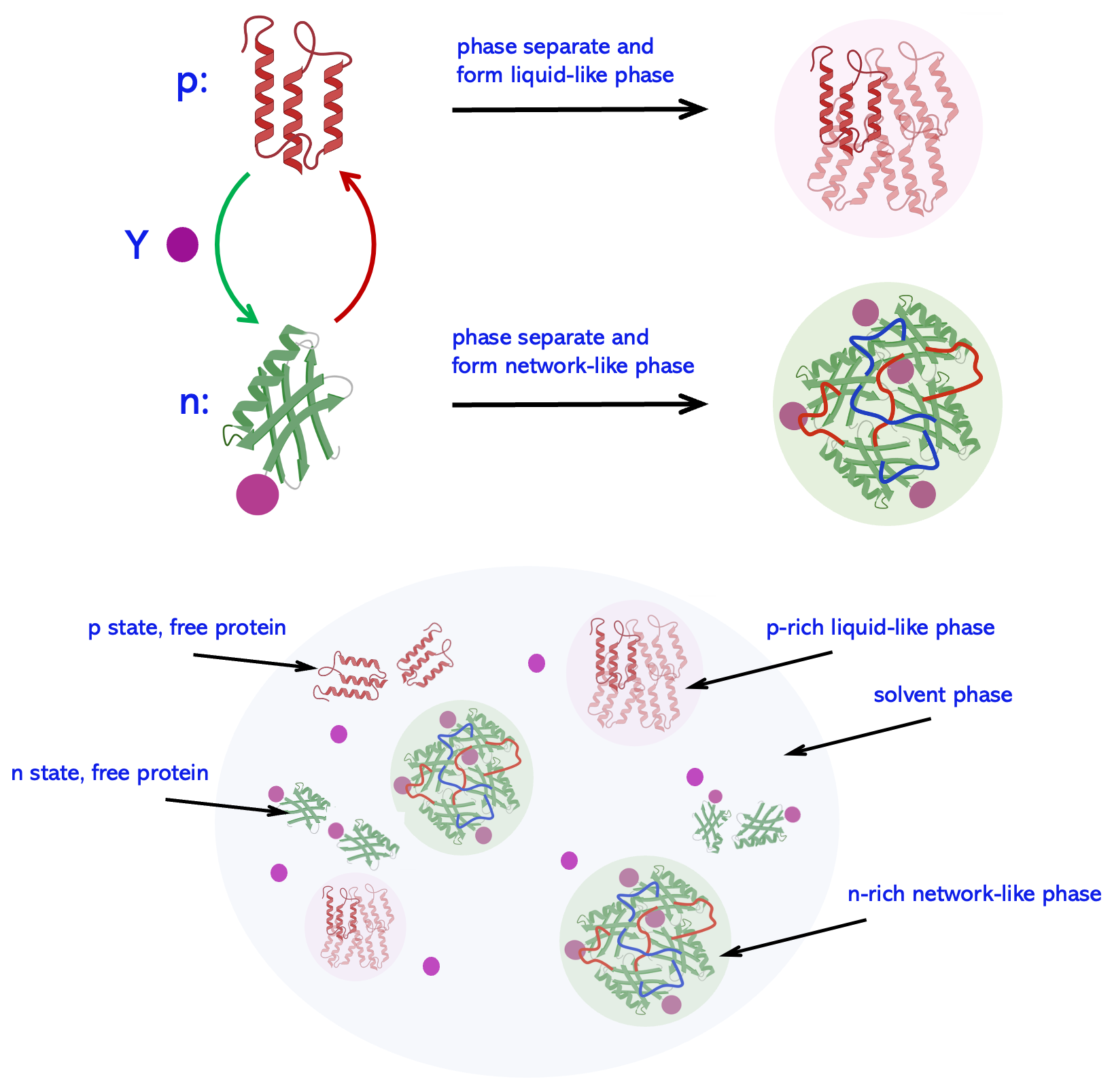}
\caption{Schematic of phase separation of the liquid-like $p-$rich and network-like $n-$rich phases.}
\label{fig:schematic}
\end{figure}

We consider that protein molecules, when they interact with another molecule or ion $Y$ in the surrounding phase, reversibly switch between two structurally or functionally distinct conformational states, denoted as $p$ and $n$, while retaining their chemical identity. 
\begin{equation}
    p \, + \, Y \xrightleftharpoons[k_{n-p}]{k_{p-n}} n
    \label{eq:rev_conv}
\end{equation}
We further assume that when protein molecules switch from $p$ state to $n$ state, such changes expose the cross-linking domains of the protein molecules, thereby increasing the propensity for dynamically cross-linked network formation. Protein molecules in both states phase separate. We consider that  the $p$-rich phase is liquid-like and the $n$-rich phase is network-like (Fig.~\ref{fig:schematic}). Therefore, the solution is a three phase mixture: $p$-rich phase, $n$-rich phase and the solvent.

Before we proceed further, let us formulate the free energy functional for this triphasic mixture. We denote $\phi_p$, $\phi_n$ and $\phi_s$ as  the volume fraction of the  $p$-rich phase, $n$-rich phase, and solvent phase, respectively.  The total balance must satisfy $(\phi_p + \phi_n + \phi_s) = 1$.
The following is the total Helmholtz free energy of this system due to the mixing free energy interactions and the free energy of the deformable protein network due to deformation gradient $\mathbf{F}$:
\begin{equation}
    \mathcal{F}(\phi_p, \phi_n, \phi_s, \mathbf{F},t) \, = \, \int_\Omega \md V \, \Big(f_{\text{mix}}(\phi_p, \phi_n,\phi_s) \, + \,  f_{\text{el,tr}}(\mathbf{F},\phi_n,t)\Big), 
    \label{eq:tot_free_en}
\end{equation}
where $f_{\text{mix}}$ is the entropic and enthalpic contribution to mixing. Following Flory–Huggins regular solution theory and Cahn–Hilliard correction for interfacial energy  \citep{doi2013soft}, 
\begin{align}
    f_{\text{mix}}(\phi_p, \phi_n, \phi_s) \, & = \, \frac{k_{\text B}T}{v} \, \Big(\phi_p \, {\rm ln} \,  \phi_p \, + \phi_n \, {\rm ln} \,  \phi_n \, + \, \phi_s \, {\rm ln} \,  \phi_s \, \nonumber \\
    & + \, \chi_{pn} \, \phi_p \, \phi_n \, + \, \chi_{ps} \, \phi_p \, \phi_s \, + \, \chi_{ns} \, \phi_n \, \phi_s \, \nonumber \\
    & + \, \frac{\kappa}{2} \, \Big[(\bm{\nabla} \, \phi_p)^2 \, + \, (\bm{\nabla} \, \phi_n)^2 \Big]  \Big),
    \label{eq:free_en_mix}
\end{align}
where $k_{\text B}$ is the Boltzmann constant, $T$ is the temperature, and $v$ is the molecular volume of the solvent. Here $\chi_{pn}$, $\chi_{ps}$, and $\chi_{ns}$ are the parameters that account for the enthalpic interactions between protein molecules in $p$ state--protein molecules in $n$ state, protein molecules in $p$ state--solvent molecules, and protein molecules in $n$ state--solvent molecules, respectively, and $\kappa$ is the interfacial coefficient. In (\ref{eq:tot_free_en}), $f_{\text{el,tr}}(\mathbf{F},\phi_n,t)$ is the contribution of free energy due to the deformation of protein networks in the $n$-rich phase that can occur during phase separation. Next, we formulate $f_{\text{el,tr}}(\mathbf{F},\phi_n,t)$ for protein networks. 

\subsection{Dynamic cross-linking and constitutive for transient network}
\label{sec:mod_2}

We assume that the protein network is transient in nature due to dynamic cross-linking events. The cross-links form and break in response to deformation during phase separation. To avoid confusion, we state that the cross-links mentioned here do not represent a new molecule in the system. Rather, they may be treated as ``sticker" segments of protein molecules with intrinsically disordered regions (IDRs). Consider that the total number of cross-links is $N_{\text{tot}} \equiv N_{\text{tot}}(\phi_n (\mathbf{x},t))$, of which the number of bound cross-links at time $t$ is $N_{b}(\phi_n(\mathbf{x},t))$, and therefore the number of unbound cross-links is $N_u$. The binding rate ($k_{\text{on}}$) of one cross-link depends on two time scales, and we assume that it remains unaffected by unbinding events of bound cross-links. In a binding event, the one end of a cross-link has to reach a binding region via diffusion, followed by reaction that accounts for successful binding. Therefore, $k_{\text{on}}$ can be approximated as   
\begin{equation}
    \frac{1}{k_{\text{on}}} \, = \, \frac{1}{k_{\text{diff}}} \, + \, \frac{1}{k_b},
    \label{eq:binding_rate}
\end{equation}
where $1/k_{\text{diff}}$ corresponds to the diffusion time scale, $k_b = k^0_b \, {\rm exp} \, (-U_b/(k_{\text{B}}T))$, and $U_b$ is the energy barrier for binding. The unbinding rate ($k_{\text{off}}$) of a cross-link follows force-dependent kinetics, where the force $f_n$ is due to the deformation in the network at time $t$, and therefore
\begin{equation}
    k_{\text{off}} (\mathbf{x},t)\, = \, k_{\text{off}}^0 \, \,  {\rm \exp} \, \Big(-\frac{U_u}{k_{\text B} T}\Big) \, \,  {\rm \exp} \,\Big(\frac{f_n (\mathbf{x}, t) \, \delta_n}{k_{\text{B}} T} \Big), 
    \label{eq:unbinding_rate_a}
\end{equation}
where $U_u$ is the energy barrier for unbinding in the absence of force, and $\delta_n$ is the size of a cross-link. Assuming affine network and uniaxial deformation  \citep{rubinstein2002elasticity}, 
\begin{equation}
    f_n  \, \delta_n \, \sim \,  \frac{3 \,  k_{\text B}T}{2 \, \delta_n^2} \, \frac{ \delta_n^2 \, (\lambda^2 \, - \, 1)}{3} \, = \, \frac{k_{\text B}T}{2} \, \Big|\lambda(x,t)^2 \, - \, 1\Big|,
    \label{eq:fn}
\end{equation}
where $\lambda$ is the elongation ratio in the $x$ direction. However, in a transient network, all cross-links do not experience the same stretch corresponding to the reference configuration. Newly formed cross-links formed at time $t^\prime$ are stress-free when they form, whereas old surviving cross-links experience stretch relative to initial configuration. To account for the former effect, we use 
\begin{equation}
    f_n (x,t;t^\prime) \, \delta_n \, \sim  \, \frac{k_{\text B}T}{2} \, \Big|\frac{\lambda^2(x,t)}{\lambda^2(x,t^\prime)} \, - \, 1\Big|,
    \label{eq:fn_1}
\end{equation}
where $\lambda(x,t)/\lambda(x,t^\prime)$ corresponds to the cohort-relative stretch for uniaxial deformation, and therefore,
\begin{equation}
    k_{\mathrm{off}}(x,t;t') \, = \, k_{\mathrm{off}}^0
    \, \exp \, \left(
    - \, \frac{U_u}{k_{\mathrm B}T}
    \right) \, 
    \exp \, \Bigg(
    \frac{1}{2} \, \Bigg|\frac{\lambda^2(x,t)}{\lambda^2(x,t^\prime)}
     - 1 \Bigg|\Bigg).
    \label{eq:unbinding_rate}
\end{equation}
For uniaxial stretching, $\lambda$ is governed by
\begin{equation}
    \lambda(x,t) \, \phi_n (x,t) \, = \phi_n^0 (x),
    \label{eq:lambda}
\end{equation}
for an incompressible mixture \citep{paulin2026dynamics}. Here $\phi_n^0$ is the  volume fraction of
a homogeneous, undeformed network in a relaxed state and $\phi_n$ is the volume fraction of the network in a deformed state. Here, (\ref{eq:fn}) is proposed in a manner in which the stretching contribution to $k_{\text{off}}$ (see (\ref{eq:unbinding_rate})) is absent due to $\lambda = 1$ (see (\ref{eq:lambda})) at time $t = 0$. 

Now, we can write the balance equation for bound cross-links $N_b$ at location $x$ as
\begin{equation}
    \frac{\md N_b}{\md t} \, = \, -\, k_{\text{off}} \, N_b \, + \, k_{\text{on}} \, N_u,
    \label{eq:crosslink_bal}
\end{equation}
where $N_u = N_{\text{tot}}(\phi_n(x,t)) - N_b (\phi_n(x,t))$. We assume that 
\begin{equation}
    N_{\text{tot}}(\phi_n(x,t)) \, \sim \, \alpha \, \phi_n(x,t),
    \label{eq:ntot}
\end{equation}
where $\alpha$ is a material parameter related to the network. Multiplying both sides of (\ref{eq:crosslink_bal}) with ${\rm exp}\,(\int_0^t \, k_{\text{off}}(x,t^\prime)\, \md t^\prime)$ and integrating with an initial condition $N_b(x, t = 0) = N_b^0 (x)$, the solution of (\ref{eq:crosslink_bal}) is 
\begin{equation}
    N_b(x,t) \, = \, \Bigg [A(t;0) \, \, + \, \, \int_0^t \, \md t^\prime \, B(x,t;t^\prime) \Bigg] \, N_b^0 (x), 
    \label{eq:crosslink_bal_sol}
\end{equation}
where $A(t)$ and $B(x,t;t^\prime)$ are
\begin{align}
    A(t;0) \, & = \, \Bigg[{\rm exp} \Big(-\int_0^t \, \md s \, k_{\text{off}}(s;0) \Big)\Bigg], 
    \label{eq:A_B1}\\ 
    B(x, t; t^\prime) \, & = \,   k_{\text{on}} \, \frac{N_u(x,t^\prime)}{N_b^0(x)} \, \, \Bigg[{\rm exp} \Big(-\int_{t^\prime}^t \, \md s \, k_{\text{off}}(s;t^\prime) \Big)\Bigg] \nonumber \\ & = \, 
    k_{\text{on}} \, \frac{\Big(\alpha \, \phi_n (x,t^\prime) \, - \, N_b(x,t^\prime)\Big)}{N_b^0(x)} \, \, \Bigg[{\rm exp} \Big(-\int_{t^\prime}^t \, \md s \, k_{\text{off}}(s;t^\prime) \Big)\Bigg]
    \label{eq:A_B2}
\end{align}
Here, the first term inside the square bracket on the right side of (\ref{eq:crosslink_bal_sol}) is due to the survivability of the cross-links initially bound up to time $t$,  and the second term for the survivability of the newly formed cross-links (that forms at time $t^\prime$) up to time $t$. We assume that the network is at equilibrium  in the beginning at $t =  0$  and, therefore, using (\ref{eq:crosslink_bal}) and (\ref{eq:ntot}), it can be deduced that  
\begin{equation}
    N_b^0(x) \, = \, \alpha  \, \phi_n^0(x) \, \frac{k_{\text{on}}}{k_{\text{off}}(t=0) \, + \, k_{\text{on}}}.
    \label{eq:nb0}
\end{equation}

We next use this former formulation on cross-linking dynamics to formulate the free energy of a transient network. To do that, we begin with an elastic rubbery network with permanent cross-links under deformation.  For reference configuration $\mathbf{X}$ at time $t = 0$ and current configuration $\mathbf{x}$ at time $t$, the deformation gradient tensor is defined as $\mathbf{F}(t;0) = \frac{\partial \mathbf{x}(t)}{\partial \mathbf{X}}$.  
We assume that there are $N_b^0$ number of bound cross-links at $t = 0$. If the cross-links are permanent, for an affine deformation at time $t$ with Green-Lagrange strain  $\mathbf{E}(t; 0) \, = \, \frac{1}{2} \, (\text{tr} [ \mathbf{F}^{\text T}(t; 0) \, \,  \mathbf{F} (t; 0) ] \, - \,  3 )$, the elastic energy density is \citep{doi2013soft} 
\begin{equation}
    f_{\text{el}}(t; 0) \, = \, \frac{1}{2} \, G \, \, \Big(\text{tr} \Big[ \mathbf{F}^{\text T}(t; 0)  \, \,  \mathbf{F} (t; 0) \Big] \, - \, 3 \Big); \, \, \, \, \, \, G \sim (k_{\text B}T/v) \, N_b^0, 
    \label{eq:rub}
\end{equation}
where the shear modulus $G$ is linearly proportional to $N_b^0$. However, since the actual protein network considered here is transient, the modified expression of $f_{\text{el}}(t; 0)$ would be due to the contribution from dynamic cross-linking, i.e. due to both terms inside the square bracket on the right side of (\ref{eq:crosslink_bal_sol}). The modified contribution to the elastic energy density from the initially bound cross-links  is (using (\ref{eq:crosslink_bal_sol}) and (\ref{eq:rub})) 
\begin{equation}
   A(t;0) \,  \, f_{\text{el}}(t; 0),   \label{eq:rub1}
\end{equation} 
and due to the newly formed cross-links at time $t^\prime$ that can survive  up to time $t$, the contribution is (using (\ref{eq:crosslink_bal_sol}) and (\ref{eq:rub})) 
\begin{equation}
  \int_0^t \, \md t^\prime \, B(t;t^\prime) \, \, f_{\text{el}}(t; t^\prime).   \label{eq:rub2}
\end{equation} 
Therefore, the final form    $f_{\text{el,tr}}$ for a transient elastic network would be 
\begin{align}
     f_{\text{el,tr}}(\mathbf{F}, \phi_n,t) \,  =  \, A(t;0) \, \, f_{\text{el}}(t; 0) \,  + \, \, \int_0^t \, \md t^\prime \, B(t;t^\prime) \, \, f_{\text{el}}(t; t^\prime).
     \label{eq:elastic_free_en}
\end{align}
A similar expression for $f_{\text{el,tr}}$ has been obtained in another work \cite{meng2016stress} for a transient polymer network. Note that $f_{\text{el}}(t;t^\prime)$ in (\ref{eq:rub2}) and (\ref{eq:elastic_free_en}) is due to the deformation $\mathbf{F}(t; t^\prime)$ in the period $t^\prime$ to $t$ contributed from the newly formed cross-links at time $t^\prime$. Using multiplicative decomposition, it can be shown that 
\begin{equation}
    \mathbf{F}(t; t^\prime) \, = \, \mathbf{F}(t;0) \, \mathbf{F}^{-1}(t^\prime;0).
    \label{eq:def_cr_link}
\end{equation}
The first Piola-Kirchhoff stress $\bm{P}_{\text{el,tr}}$ and Cauchy stress $\bm{\sigma}_{\text{el,tr}}$ are defined as
\begin{equation}
  \bm{P}_{\text{el,tr}} \, = \, \frac{\delta \mathcal{F}}{\delta \mathbf{F}} \, - \, p\, \frac{\delta J} {\delta \mathbf{F}}, \; \; \; \bm{\sigma}_{\text{el,tr}} \, = \, \frac{1}{J} \, \bm{P}_{\text{el,tr}} \, 
  \bm{F}^{T},    
  \label{eq:stress_tensor}
\end{equation}
where $p$ is the pressure, called the Lagrange multiplier, to enforce incompressibility.

\subsection{Balance equations}
\label{sec:mod_3}

The component balance equations for three phases are
\begin{align}
    \frac{\partial \phi_p}{ \partial t} \, + \, \bm{\nabla} \cdot (\phi_p \, \mathbf{v}_p) \, & = \, - \mathcal{R},  \nonumber\\
    \frac{\partial \phi_n}{ \partial t} \, + \, \bm{\nabla} \cdot (\phi_n \, \mathbf{v}_n) \, & = \,   \mathcal{R}, \nonumber\\
    \frac{\partial \phi_s}{ \partial t} \, + \, \bm{\nabla} \cdot (\phi_s \, \mathbf{v}_s) \, & = \, 0,
    \label{eq:massbal}
\end{align}
where  $\mathbf{v}_p$, $\mathbf{v}_n$, and $\mathbf{v}_s$ are the velocities of the $p$-rich phase, $n$-rich phase, and solvent phase, respectively. We set the choice of $\mathcal{R}$ as (see (\ref{eq:rev_conv}))
\begin{equation}
    \mathcal{R} \, = \, k_{p-n} \, \phi_p \, - \, k_{n-p} \, \phi_n.
    \label{eq:reaction_type2}
\end{equation}
For simplicity, in this work, we treat the forward reaction in (\ref{eq:rev_conv}) as a first order reaction, assuming that $Y$ is abundant and its concentration  effectively remains uniform and constant across the domain; $k_{p-n}$ and $k_{n-p}$ are rate constants related to conversion reactions (see (\ref{eq:rev_conv})). Here,  $\mu_p$ and $\mu_n$ are the chemical potentials of the $p$-rich phase and $n$-rich phase, respectively:
\begin{align}
   \mu_p \, = \,   \frac{\delta \mathcal{F}}{\delta \phi_p} \,  = & \, \frac{k_{\text B} T}{v}\left[1 \, + \, {\rm ln}(\phi_p) \, + \, \chi_{pn} \, \phi_n \, + \, \chi_{ps} \, \phi_s \, - \, \kappa \, \bm{\nabla}^2 \, \phi_p \right];  \\
   \mu_n \, = \,  \frac{\delta \mathcal{F}}{\delta \phi_n}  \,   = & \,  \frac{k_{\text B} T}{v}\left[1 \, + \, {\rm ln}(\phi_n) \, + \, \chi_{pn} \, \phi_p \, + \, \chi_{ns} \, \phi_s \, - \, \kappa \, \bm{\nabla}^2 \, \phi_n \right] \, +   \, \frac{\partial f_{\text{el,tr}}}{\partial \phi_n}.    
   \label{eq:chem_pot}
\end{align}
The last term  in (\ref{eq:chem_pot}) is due to $f_{\text{el,tr}}(\mathbf{F}, \phi_n,t)$ which is a function of $\phi_n$ (see (\ref{eq:tot_free_en}) and (\ref{eq:elastic_free_en})).  
The summation of the three equations in (\ref{eq:massbal}) results in
\begin{equation}
    \bm{\nabla} \cdot (\phi_p \, \mathbf{v}_p \, + \, \phi_n \, \mathbf{v}_n \, + \, \phi_s \, \mathbf{v}_s) \, = \, 0 
    \label{eq:bal},
\end{equation}
as $(\phi_p + \phi_n + \phi_s) = 1$. We define the mixture velocity as $\mathbf{v} \, = \, (\phi_p \, \mathbf{v}_p \, + \, \phi_n \, \mathbf{v}_n \, + \, \phi_s \, \mathbf{v}_s)$. If there is no net flow of the mixture in a system with periodic boundary conditions, $\mathbf{v}$ can be set to 0. In this work, we set $\mathbf{v} = 0$. Next, we will discuss force balance on the $p$-rich and $n$-rich phases. 

We treat the $p$-rich phase
as a simple liquid and assume that internally generated viscous stresses relax rapidly, i.e., the $p$-rich phase cannot sustain long-lived shear stresses. Furthermore, we assume that momentum dissipation is dominated by interphase friction, that arises from the relative motion of the $p$-rich phase with respect to the barycentric/background velocity field, rather than viscous dissipation within the $p$-rich phase, i.e. $\nabla \cdot \bm{\sigma}_p \ll \xi_p \, (\mathbf{v} \, - \,  \mathbf{v}_p)$. Therefore, the thermodynamic driving force due to the chemical potential gradient is balanced by the frictional drag force due to the relative motion of the $p$-rich phase with respect to the background flow.  Neglecting inertial effects in the creeping flow limit, the force balance for the $p$-rich phase  reads as 
\begin{equation}
    -\,\phi_p \, v \, \bm{\nabla} \mu_p  \, + \, \xi_p \, (\mathbf{v} \, - \,  \mathbf{v}_p) \,  =  \, 0
    \label{eq:forcebal_p},
\end{equation} 
where $\xi_p$ is the friction (or drag) coefficient.

As the $n$-rich phase is dynamically cross-linked elastic network, the deformation of the network during phase separation can generate internal stresses, consequently restoring forces. Therefore, the thermodynamic driving force due to the chemical potential gradient must be balanced by the restoring forces and the frictional drag that arises from the relative motion of the network phase with respect to the background flow. Neglecting inertial effects, the force balance for the $n$-rich network phase is \citep{onuki1999spinodal}
\begin{equation}
    -\,\phi_n \, v \, \bm{\nabla} \mu_n  \, + \, v \, \bm{\nabla} \cdot \bm{\sigma}_{\text{el,tr}} \, + \, \xi_n \, (\mathbf{v} \, - \,  \mathbf{v}_n) \,   \, =  \, 0,
    \label{eq:forcebal_n}
\end{equation}
where $\xi_n$ is the friction coefficient related to the movement of the $n$-rich phase in the mixture. Setting $\mathbf{v} = 0$, and substituting (\ref{eq:forcebal_p}) and (\ref{eq:forcebal_n}) into the first two of (\ref{eq:massbal}),
\begin{align}
    \frac{\partial \phi_p}{ \partial t} \, & = \, \bm{\nabla} \cdot \Big(\frac{D_p \, v}{k_{\text B}T}\, \phi_p^2 \, \bm{\nabla}\, \mu_p \Big) \, - \, \mathcal{R}, 
    \label{eq:massbal_p}\\
    \frac{\partial \phi_n}{ \partial t} \, & = \, \bm{\nabla} \cdot \Big(\frac{D_n \, v}{k_{\text B}T}\, \phi_n^2 \, \bm{\nabla}\, \mu_n \Big) \, - \, \bm{\nabla} \cdot \Big(\frac{D_n \, v}{k_{\text B}T} \, \phi_n \, \bm{\nabla} \cdot  \bm{\sigma}_{\text{el,tr}} \Big)  \, + \, \mathcal{R},
    \label{eq:massbal_n}
\end{align}
where $D_p = \frac{k_{\text B}T}{\xi_p}$ and $D_n = \frac{k_{\text B}T}{\xi_n}$.

\subsection{Equations for 1D system}
\label{sec:mod_4}

We consider a one dimensional system and uniaxial stretching of the network that can occur during phase separation. Using (\ref{eq:chem_pot}), (\ref{eq:massbal_p}) and (\ref{eq:massbal_n}) are reduced to the following:
\begin{align}
\frac{\partial \phi_p}{\partial t} & \, = \,  \frac{\partial}{\partial x}
\left[
M_p\,
\phi_p^2 \,
\frac{\partial}{\partial x} \Big(\frac{\delta \mathcal{F}}{\delta \phi_p}\Big)
\right] \, 
- \, \mathcal{R},
\label{eq:1d_mup}
\\
\frac{\partial \phi_n}{\partial t}
& \, = \,
\frac{\partial}{\partial x} \, 
\left[
M_n\,
\phi_n^2 \,
\frac{\partial}{\partial x} \Big(\frac{\delta \mathcal{F}}{\delta \phi_n}\Big)
\right] \, 
- \, 
\frac{\partial}{\partial x}
\left[
M_n \,
\phi_n\,
\frac{\partial}{\partial x} (\sigma_{\text {el,tr}}|_{xx})
\right] \, + \, 
\mathcal{R}.
\label{eq:1d_mun}
\end{align}
where $M_p = (D_p\, v)/(k_{\text B}T)$ and $M_n = (D_n\, v)/(k_{\text B}T)$. 

To solve the above set of equations, we require $\sigma_{\text {el,tr}}|_{xx}$, which we will obtain in the following way. Consider the network under incompressible uniaxial stretch in a 1D configuration. For a stretch in the $x-$ direction with an elongation ratio  $\lambda (x,t)$ that can vary spatially and temporally, $\mathbf{F}(t;0)$ can be expressed in terms of $\lambda (x,t)$ as,
\begin{equation}
    F(t;0) \, = \, 
\begin{pmatrix}
\lambda(t) & 0 & 0 \\
0 & \lambda^{-1/2}(t) & 0 \\
0 & 0 & \lambda^{-1/2}(t)
\end{pmatrix}.
 \label{eq:def_elong1}
\end{equation}
Using (\ref{eq:rub}), (\ref{eq:def_cr_link}), and (\ref{eq:def_elong1}), $f_{\text{el}}(t;0)$ and $f_{\text{el}}(t;t^\prime)$ for this 1D configuration can be written as
\begin{align}
    f_{\text{el}}(t;0) \, & = \, \frac{1}{2} \, G \, \Big(\lambda^2(x,t) \, + \, \frac{2}{\lambda(x,t)} - 3
    \Big), 
    \nonumber \\
    f_{\text{el}}(t;t^\prime) \, & = \, \frac{1}{2} \, G \, \Bigg( \frac{\lambda^2(x,t)}{\lambda^2(x,t^\prime)} \,  
    \, + \, 2 \, \frac{\lambda(x,t^\prime)}{\lambda(x,t)} - 3 
    \Bigg).
    \label{eq:free_en_elong1}
\end{align}
Substituting (\ref{eq:free_en_elong1}) into (\ref{eq:elastic_free_en}), we obtain  
\begin{align}
     f_{\text{el,tr}}(\mathbf{F},\phi_n,t) \,  & =  \, \frac{1}{2} \, G \, \, A(t;0) \, \, \Big(\lambda^2(x,t) \, + \, \frac{2}{\lambda(x,t)} - 3 
     \Big) \,  \nonumber \\ &+ \, \frac{1}{2} \, G \, \int_0^t \, \md t^\prime \, B(x,t;t^\prime) \, \, \Bigg( \frac{\lambda^2(x,t)}{\lambda^2(x,t^\prime)} \, + \, 2 \, \frac{\lambda(x,t^\prime)}{\lambda(x, t)} - 3 
     \Bigg).
     \label{eq:elastic_free_en_subs}
\end{align}
Therefore, it can be shown that
\begin{equation}
\frac{\partial f_{\text{el,tr}}}{\partial \phi_n} \, \approx \, -\frac{G}{\phi_n(t)} \, \Bigg[ A(t;0) \, \Bigg( \lambda^2(x,t) \, - \, \frac{1}{\lambda(x,t)} \Bigg) \, + \, 
\int_0^t \md t^\prime B(t;t^\prime)
\Bigg( \frac{\lambda^2(x,t)}{\lambda^2(x,t^\prime)} \,  - \,  \frac{\lambda(x,t^\prime)}{\lambda(x,t)}
\Bigg)
\Bigg].
\label{eq:free_en_el}
\end{equation}
For incompressible case, $J = \det \mathbf{F} = 1$.  Using (\ref{eq:stress_tensor}) and the incompressibility constraint, and eliminating the pressure by imposing vanishing transverse stress, $\sigma_{\text{el,tr}}|_{xx}$ would be,
\begin{align}
    \sigma_{\text{el,tr}}|_{xx} \,  \equiv \, \sigma_{xx}\, =  & \, G \, \, A(t;0) \,  \Bigg[\lambda^2(x,t)
    \, - \, \frac{1}{\lambda(x,t)}\Bigg] \, \nonumber \\
    & + \,  \, G \, \int_0^t \, \md t^\prime \, B(x, t;t^\prime) \, \, \Bigg[ 
    \frac{\lambda^2(x,t)}{\lambda^2(x,t^\prime)} \, - \, \frac{\lambda(x,t^\prime)}{\lambda(x,t)} 
    \Bigg].
     \label{eq:stress_xx}
\end{align}

The summary of the equations are shown in table~\ref{tab:mod_eqns}.

\begin{table}[hbtp]
\centering
\begin{tabular}{c}
\hline 
1D system\\
\hline 
$\frac{\partial \phi_p}{\partial t}  \, = \,  \frac{\partial}{\partial x}
\left[
M_p\,
\phi_p^2 \,
\frac{\partial}{\partial x} \Big(\frac{\delta \mathcal{F}}{\delta \phi_p}\Big)
\right] \, 
- \, \mathcal{R}$
\\
$\frac{\partial \phi_n}{\partial t}
 \, = \,
\frac{\partial}{\partial x} \, 
\left[
M_n\,
\phi_n^2 \,
\frac{\partial}{\partial x} \Big(\frac{\delta \mathcal{F}}{\delta \phi_n}\Big)
\right] \, 
- \, 
\frac{\partial}{\partial x}
\left[
M_n \,
\phi_n\,
\frac{\partial}{\partial x} (\sigma_{\text {el,tr}}|_{xx})
\right] \, + \, 
\mathcal{R}$ \\
$\mathcal{R} \, = \, k_{p-n} \, \phi_p \, - \, k_{n-p} \, \phi_n$ \\
$\phi_s \, = \, 1 \, - \, \phi_p \, - \, \phi_n $ \\
$M_p = (D_p\, v)/(k_{\text B}T)$ and $M_n = (D_n\, v)/(k_{\text B}T)$ \\
$\mu_p \, = \,   \frac{\delta \mathcal{F}}{\delta \phi_p} \,  =  \, \frac{k_{\text B} T}{v}\left[1 \, + \, {\rm ln}(\phi_p) \, + \, \chi_{pn} \, \phi_n \, + \, \chi_{ps} \, \phi_s \, - \, \kappa \, \bm{\nabla}^2 \, \phi_p \right]$ \\
$\mu_n \, = \,  \frac{\delta \mathcal{F}}{\delta \phi_n}  \,   =  \,  \frac{k_{\text B} T}{v}\left[1 \, + \, {\rm ln}(\phi_n) \, + \, \chi_{pn} \, \phi_p \, + \, \chi_{ns} \, \phi_s \, - \, \kappa \, \bm{\nabla}^2 \, \phi_n \right] \, +   \, \frac{\partial f_{\text{el,tr}}}{\partial \phi_n}$ \\
$f_{\text{el,tr}}(\mathbf{F},\phi_n,t) \,  =  \, \frac{1}{2} \, G \, \, A(t;0) \, \, \Big(\lambda^2(x,t) \, + \, \frac{2}{\lambda(x,t)} - 3 
     \Big) \, $\\$+ \, \frac{1}{2} \, G \, \int_0^t \, \md t^\prime \, B(x,t;t^\prime) \, \, \Bigg( \frac{\lambda^2(x,t)}{\lambda^2(x,t^\prime)} \, + \, 2 \, \frac{\lambda(x,t^\prime)}{\lambda(x, t)} - 3 
     \Bigg)$\\
$\lambda(x,t) \, \phi_n (x,t) \, = \, \phi_n^0 (x)$\\
$\frac{\partial f_{\text{el,tr}}}{\partial \phi_n} \, \approx \, -\frac{G}{\phi_n(t)} \, \Bigg[ A(t;0) \, \Bigg( \lambda^2(x,t) \, - \, \frac{1}{\lambda(x,t)} \Bigg) \, + \, 
\int_0^t \md t^\prime B(t;t^\prime)
\Bigg( \frac{\lambda^2(x,t)}{\lambda^2(x,t^\prime)} \,  - \,  \frac{\lambda(x,t^\prime)}{\lambda(x,t)}
\Bigg)
\Bigg]$ \\
$\sigma_{\text{el,tr}}|_{xx} \,   =  \, G \, \, A(t;0) \,  \Bigg[\lambda^2(x,t)
    \, - \, \frac{1}{\lambda(x,t)}\Bigg] \, + \,  \, G \, \int_0^t \, \md t^\prime \, B(x, t;t^\prime) \, \, \Bigg[ 
    \frac{\lambda^2(x,t)}{\lambda^2(x,t^\prime)} \, - \, \frac{\lambda(x,t^\prime)}{\lambda(x,t)} 
    \Bigg]$ \\
$A(t;0) \, = \, \Bigg[{\rm exp} \Big(-\int_0^t \, \md s \, k_{\text{off}}(s;0) \Big)\Bigg]$ \\
$B(x,t; t^\prime) \,  = \,   k_{\text{on}} \, \frac{\Big(\alpha \, \phi_n (x,t^\prime) \, - \, N_b(x,t^\prime)\Big)}{N_b^0(x)} \, \, \Bigg[{\rm exp} \Big(-\int_{t^\prime}^t \, \md s \, k_{\text{off}}(s;t^\prime) \Big)\Bigg]$ \\
$G \sim (k_{\text B}T/v) \, N_b^0(x)$ \\
$N_b^0(x) \, = \, \alpha  \, \phi_n^0(x) \, \frac{k_{\text{on}}}{k_{\text{off}}(t=0) \, + \, k_{\text{on}}}$ \\
$\frac{N_b(x,t)}{N_b^0(x)} \, = \, A(t;0) \, \, + \, \, \int_0^t \, \md t^\prime \, B(x,t;t^\prime) $ \\
$\frac{1}{k_{\text{on}}} \, = \, \frac{1}{k_{\text{diff}}} \, + \, \frac{1}{k_b}$ where $k_b = k^0_b \, {\rm exp} \, (-U_b/(k_{\text{B}}T))$ \\
Non-aging: $k_{\mathrm{off}}(x,t;t') \, = \, k_{\mathrm{off}}^0
    \, \exp \, \left(
    - \, \frac{U_u}{k_{\mathrm B}T}
    \right) \, 
    \exp \, \Bigg(
    \frac{1}{2} \, \Bigg|\frac{\lambda^2(x,t)}{\lambda^2(x,t^\prime)}
     - 1 \Bigg|\Bigg)$ \\
\hline
\end{tabular}
\caption{Summary of model equations.}
\label{tab:mod_eqns}
\end{table}

\subsection{Aging of network phase} 
\label{sec:mod_5}

The framework discussed until now is suitable for a non-aging network. We hypothesize that the transition from liquid-like condensate to aging condensate is not a single pathway. Protein molecules that form liquid-like phases can form network-like phases when the biochemical environment changes, where the latter can alter binding/unbinding interactions. There can be two possibilities for a further transition. In one case, network phases remain transient through dynamic cross-linking, implying non-aging. In the second case, condensates age as time elapses and the mechanical response of the network changes with time. Aging of a condensate environment may be due to many reasons, including continuous conformational change of proteins, increased local ordering, network densification via stabilization of cross-links, solvent expulsion, and insufficiency of molecules like molecular chaperones and ATP molecules inside the condensate environment \citep{Yoo2022Chaperones,linsenmeier2022dynamic,garaizar2022aging,Alshareedah2024, biswas2024molecular}. We hypothesize that all these former factors that contribute to condensate aging directly or indirectly stabilize the cross-links of the network progressively over time. In a phenomenological manner, we propose the modified unbinding rate of cross-links of a network at time $t$ that is aged for a time $t_s$ as (see  \ref{sec:App_a} for details)
\begin{equation}
    k_{\mathrm{off}}(t_s, t;t^\prime) \, = \, k_u \, 
    k_\lambda \, \Big( 1 \, + \, \frac{t \, - \, t_s}{\tau_a} \Big)^{-a}, 
\label{eq:unbinding_rate_aging}
\end{equation}
where $\tau_a$ is the characteristic time scale of aging and $a > 0$ governs the strength of aging. Here $a = 0$ corresponds to a non-aging network, while the network ages for the chosen value of $a = 1.5$.  The dimensionless number $\tilde{\tau}_a = k_u \tau_a$ governs the relative aging rate. There are other modeling approaches to model $\kon$ and $\koff$ \citep{zaccone2013colloidal,debnath2026multi}; the current work follows a simple phenomenological description for $\kon$ and $\koff$ to reduce complexity of the overall model for the time being. If $\tau_a \gg (1/k_u)$, aging is slower compared to the initial renewal time of cross-links. In the former, many cross-link formation and breakage events occur before a network ages appreciably, in which the fluid-like nature of a network slowly reduces. In contrast, if $\tau_a \sim (1/k_u)$, aging and cross-linking events occur on comparable timescales. For the case $\tau_a \sim (1/k_u)$, aging progresses rapidly before a network retains its fluidic behavior through cross-link formation and breakage. Consequently, the population of long-lived older cross-links  increases rapidly over time in the case $\tau_a \sim (1/k_u)$ in a rapidly aging network.  

\section{Results}
\label{sec:res}
De-linking phase separation, we next investigate how cross-linking dynamics governs the mechanical properties
of network-like phases for both non-aging and aging cases.

\subsection{Oscillatory strain and Maxwell fluid} 
\label{sec:res1}

\begin{figure}
\centering
\includegraphics[width=\linewidth,height=0.68\textheight,keepaspectratio]{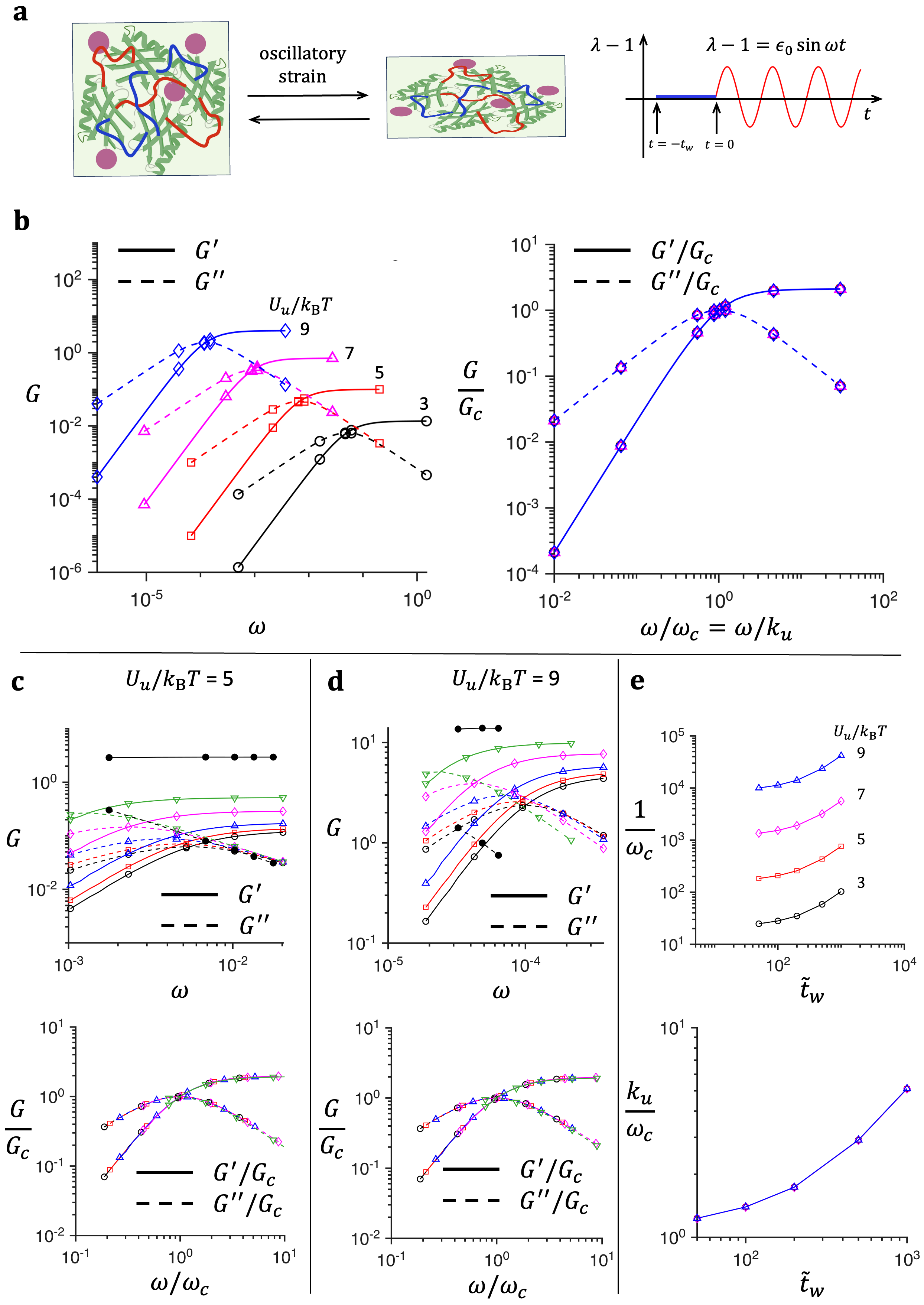}
\caption{\textbf{Oscillatory strain.} (a) Schematic of oscillatory strain experiment. (b) Results for non-aging network: with frequency $\omega$ and its scaled value $\omega/\omega_c$, the variation of dimensional ($G^\prime,  \, G^{\prime \prime}$) and dimensionless moduli  ($G^\prime/G_c,  \, G^{\prime \prime}/G_c$), respectively; $G_c$ is the modulus at the cross-over frequency $\omega_c$. (c)--(e) Results for aging network. The variations of ($G^\prime,  \, G^{\prime \prime}$) and ($G^\prime/G_c,  \, G^{\prime \prime}/G_c$) with $\omega$ and $\omega/\omega_c$, respectively, for cases $U_u/\kb T = 5$ (c) and $U_u/\kb T = 9$ (d). In (c) and (d), the data are for different values of the scaled waiting time $\tilde{t}_w$:  $\circ$, $\tilde{t}_w = 50$; $\square$, $\tilde{t}_w = 100$; $\triangle$, $\tilde{t}_w = 200$; $\diamond$, $\tilde{t}_w = 500$; $\triangledown$, $\tilde{t}_w = 1000$; $\bullet$, $\tilde{t}_w = 5000$. In (e), the variation of the inverse of cross-over frequencies $1/\omega_c$ and their scaled values $k_u/\omega_c$ are shown with $\tilde{t}_w$. For all results, $U_b/\kb T = 10$, $\alpha = 5$ and $\phi_n^0 =1$. For aging case, $\tilde{\tau}_a = k_u \, \tau_a = 500$ and $a = 1.5$. }
\label{fig:os_strain}
\end{figure}

We perform stress relaxation and oscillatory strain tests (see  \ref{sec:App_b1} and  \ref{sec:App_b2} for details) of the networks after they are aged for a waiting time period $t_w$. Here 
$\lambda(t)= 1$ in the period $-t_w < t < 0$, implying a stationary state followed by oscillations with $\lambda = 1  +  \epsilon_0  \sin \omega t$ when $t\ge 0$ (Fig.~\ref{fig:os_strain}a). For the non-aging case, the storage $G^\prime$ and loss $G^{\prime \prime}$ moduli can be obtained using Eq.~\ref{eq:stress_xx} \citep{ferry,mewis2012colloidal} (see  \ref{sec:App_b2}). The trends of $G^\prime$ and $G^{\prime \prime}$ in Fig.~\ref{fig:os_strain}b indicate the behavior of a visco-elastic Maxwell fluid. As expected, with an increase in the unbinding energy barrier $U_u/\kb T$, the cross-over frequency shifts to the left for a fixed value of the binding energy barrier $U_u/\kb T$ (Fig.~\ref{fig:os_strain}b). The former infers the formation of a relatively stronger network with a reduction in fluid-like nature for a higher value of $U_u/\kb T$ compared to a lower value of $U_u/\kb T$. However, the scaled trends of moduli for all $U_u/\kb T$ collapse, implying that the relaxation time of non-aging networks, which is inverse of the cross-over frequency $\omega_c$, is none other than the time-scale for force-free unbinding $1/k_u$ (see the second panel of Fig.~\ref{fig:os_strain}b), irrespective of $U_u/\kb T$. The findings in Fig.~\ref{fig:os_strain}b is further supported by a step strain test followed by stress relaxation (see  \ref{sec:App_b1} and Fig.~\ref{fig:stress_relax}a-c for details). The relaxation trends of non-aging transient networks are similar to that of a Maxwell fluid.

In the aging case, the network mechanics changes significantly depending on the waiting time $t_w$ and the speed of aging of a network (see  \ref{sec:App_b2} for details). From stress relaxation study, there are two limiting cases depending on the dimensionless characteristic timescale for aging: $\tilde{\tau}_a = k_u \tau_a \gg 1$ and $\tilde{\tau}_a = k_u \tau_a \sim 1$ (see  \ref{sec:App_b1}). Consider the limit $\tilde{\tau}_a \gg 1$. The rate of aging is slower; stress relaxes and recovery  occurs over a longer period of time (Fig.~\ref{fig:stress_relax}c,d). The stress relaxation trends for different $\tilde{t}_w = k_u t_w$ indicate that the relaxation time depends on the waiting time $t_w$, which is reflected in the oscillatory strain test also.  From oscillatory strain test, we observe that the cross-over frequency shifts to the left (Fig.~\ref{fig:os_strain}c,d) as $\tilde{t}_w = k_u t_w$ increases. For a significantly larger value of $\tilde{t}_w$, the trends of instantaneous $G^\prime$ and $G^{\prime \prime}$ do not cross (Fig.~\ref{fig:os_strain}c,d). These findings imply that a network behaves as a  Maxwell fluid during the progression of aging. However, the transient behavior of the network reduces dramatically as it gradually and significantly ages, irrespective of $U_u/\kb T$. When networks behave as an aging Maxwell fluid, their relaxation time $1/\omega_c$ strongly varies with the waiting time and $U_u/\kb T$ (Fig.~\ref{fig:os_strain}e).  The second panel of Fig.~\ref{fig:os_strain}e shows
a similar scaling of the relaxation time $1/\omega_c$ by the time-scale for force-free unbinding $1/k_u$ for the aging case that is observed in non-aging cases, irrespective of $U_u/\kb T$. The findings in Fig.~\ref{fig:os_strain}e are qualitatively similar to that reported in previous experiments \citep{jawerth2020protein}. In brief, the findings from oscillatory tests infer that a non-aging network always behaves like a Maxwell fluid, whereas an aging network shows a transition from a Maxwell fluid during progression of aging to solid-like after  significant aging.

In the the other limit, $\tilde{\tau}_a \sim 1$. In this regime, stress can relax, but with  incomplete recovery and non-zero residual stress as a function of the waiting time $t_w$ (see \ref{sec:App_b1} and Fig.~\ref{fig:stress_relax}e,f). When $\tilde{\tau}_a = k_u \tau_a = 1$, the aging time scale $\tau_a$ is set to the same value as the cross-link unbinding time scale $1/k_u$, implying an increase in the speed of aging, compared to $\tilde{\tau}_a \gg 1$. Aging progresses very rapidly with a small increase in  $\tilde{t}_w$, consequently an increase in scaled residual stress. When $\tilde{\tau}_a \sim 1$, the higher value of the scaled residual stress  close to 1 at a higher value of  $\tilde{t}_w$  indicates that the non-dissipating stress is more like elastic when the strain is held constant (Fig.~\ref{fig:stress_relax}e,f). We speculate that the aging network behaves as a visco-elastic solid-like material when it ages significantly and very rapidly. This finding leads us to perform active microrheology-based creep test to determine whether the model captures the transition from  a viscoelastic Maxwell fluid to a dynamically arrested viscoelastic Kelvin-Voigt solid.

\subsection{Active microrheology-based creep and Kelvin-Voigt solid}

\begin{figure}
\centering
\includegraphics[width=0.7\linewidth,height=0.68\textheight,keepaspectratio]{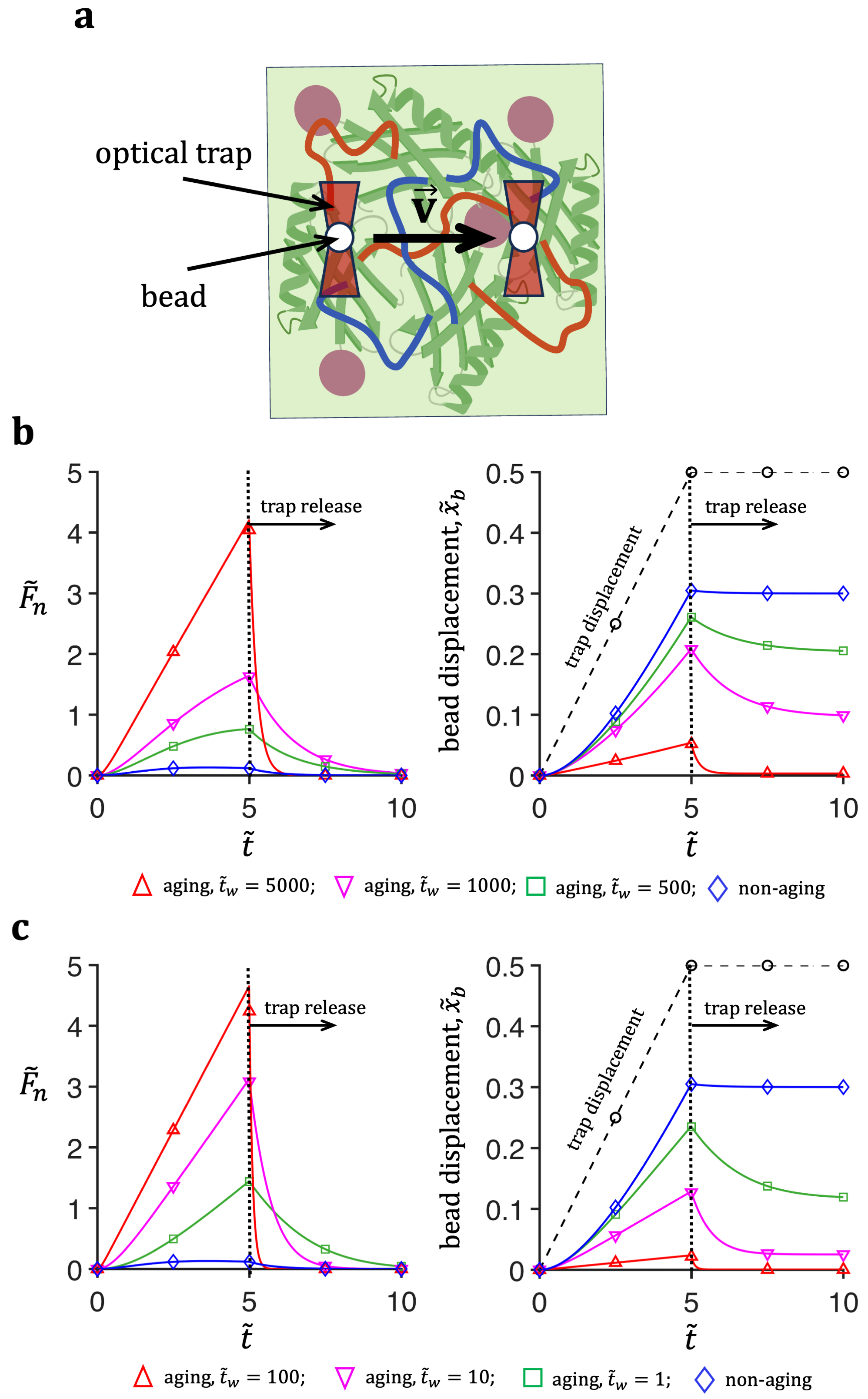}
\caption{\textbf{Creep test.} (a) Schematic of active microrheology-based creep test. Dimensionless force $\tilde F_n = F_n/(c_n G^*)$  (see details in \ref{sec:App_b3}) exerted by the network on the bead  and bead displacement $\tilde x_b = x_b/l_n$ with dimensionless time $\tilde t = k_u t$ for cases (b) $\tilde{\tau}_a = 500$ and (c) $\tilde{\tau}_a = 1$. Parameters: $U_u/\kb T = 5$, $U_b/\kb T = 10$,  $\alpha = 5$, $\phi_n^0 =1$, $a = 1.5$, $\tilde \xi_b = 20$ and $\tilde k_{\text{tr}} = 10$. }
\label{fig:ac_creep}
\end{figure}

Consider a bead embedded within a  network. The movement of an optical trap applies force on the bead to displace, and at the same time the network resists the motion of the bead (Fig.~\ref{fig:ac_creep}a, and see details in \ref{sec:App_b3}). 
The force exerted by the network on the bead before and after the trap is released, and the trends of bead trajectories after the release of the trap reveal a clear aging-induced transition from Maxwell fluid to Kelvin-Voigt solid (Fig.~\ref{fig:ac_creep}b,c). The non-aging network shows the least recoil after the trap is released. In contrast, a strong elastic recoil is exhibited by a rapidly aging network. The recovery response or the extent of the recoil increases with scaled waiting time $\tilde t_w = k_u t_w$, implying the progressive buildup of elastic memory. When the characteristic time scale for aging $\tilde \tau_a$ is larger, the structural evolution is slow, and the network retains predominantly a Maxwell fluid, unless it is aged for a long period of time $\tilde{t}_w$ (Fig.~\ref{fig:ac_creep}b). However, in rapid aging $\tilde \tau_a \sim 1$,  even a much shorter waiting period $\tilde{t}_w$ can produce elastic recoil, and the network quickly develops the features of a Kelvin-Voigt solid (Fig.~\ref{fig:ac_creep}c). The findings in Fig.~\ref{fig:ac_creep} are qualitatively similar to experimental trends reported for mutated and non-mutated A1-LCD condensate systems \citep{Alshareedah2024}. Thus, the competition between the waiting time and the characteristic time scale for aging controls the speed of transition from dissipative flow to elastic recoil for an aging network.

\subsection{Interplay between cross-linking dynamics  and molecular switching kinetics drives morphological transitions}

\begin{figure}
\centering
\includegraphics[width=\linewidth,height=0.64\textheight,keepaspectratio]{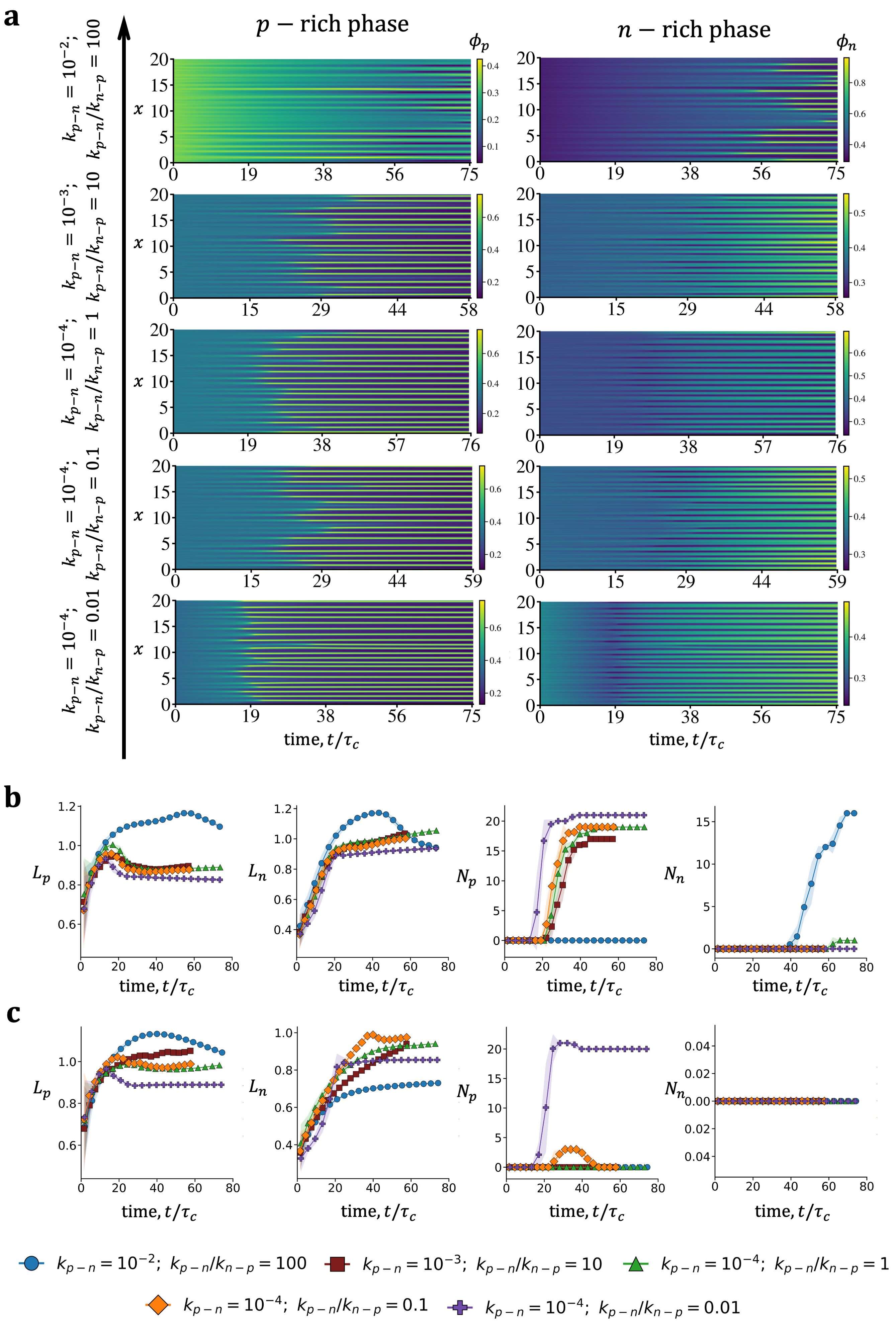}
    \caption{\textbf{Morphology.} (a) Color plots representing the spatial evolution of the $p-$rich and $n-$rich phases with time $t/\tau_c$ (where $\tau_c = (\delta t)_{\text{max}} = 10^{-5}$) during phase separation. The colors represent the values of volume fractions $\phi_p$ and $\phi_n$. The color plots in (a) are for the case $U_b/\kb T = 10$, $U_u/\kb T = 5$. In (b) and (c), the characteristic length scales $L_p$ and $L_n$ and number of domains $N_n$ and $N_n$ of the $p-$rich and $n-$rich phases, respectively are plotted over time. In (b) $U_b/\kb T = 10$, $U_u/\kb T = 5$ and in (c), $U_b/\kb T = 5$, $U_u/\kb T = 10$. These results correspond to $\alpha = 5$, $\chi_{pn} = 5$, and $\chi_{ps} = \chi_{ns} =  3$. (See \ref{sec:App_c} for details.) In (b) and (c), $N_p$ and $N_n$ represent number of phase-separating strips that reach the volume fraction 0.6 within the time frame the simulations are performed.   
}
\label{fig:morph}
\end{figure}

After the formation of a network-like phase with $\phi_n \rightarrow 1$, how a non-aging network responds to different mechanical loadings and how its mechanical features vary in aging have been discussed before. We next focus on how formation of a non-aging network-like phase through molecular switching and dynamic cross-linking impacts phase separation dynamics. The coupled set of equations in Table~\ref{tab:mod_eqns} are solved for a one-dimensional triphasic mixture with initial composition $\phi_p^0 = \phi_n^0 = 0.35$ and $\phi_s = 0.3$ (see \ref{sec:App_c} for details). When cross-link unbinding is faster than the binding time scale ($U_b/\kb T = 10$ and $U_u/\kb T = 5$), the molecular switching kinetics affect the network formation speed, particularly when the forward rate of molecular switching from $p-$ to $n-$ conformational state, i.e. $k_{p-n}/k_{n-p}$ increases (see the first panel of Fig.~\ref{fig:morph}a). With a decrease in $k_{p-n}/k_{n-p}$, the formation of the $p-$rich phase is accelerated, and the reduction in $k_{p-n}/k_{n-p}$ delays the coarsening dynamics of the $n-$rich phase. The effect of $k_{p-n}/k_{n-p}$ is also reflected in the onset of phase separation $t_{\text{onset}}$ and the duration of phase separation $(t^* - t_{\text{onset}})$ (Fig.~\ref{fig:time_morph_1}). Here $t_{\text{onset}}$ represents an average time for phase-separating  stripes of the $p-$rich and $n-$rich phases to reach volume fraction 0.4, and $(t^* - t_{\text{onset}})$ represents the duration to reach volume fraction from 0.4 to 0.6.  The findings in Fig.~\ref{fig:morph}a and Fig.~\ref{fig:time_morph_1} indicate a strong role of cross-linking dynamics in the  onset of phase separation and the speed of coarsening. 

\subsubsection*{Binding-unbinding competition regulates phase formation and coarsening}
\begin{figure}
\centering
\includegraphics[width=0.7\linewidth]{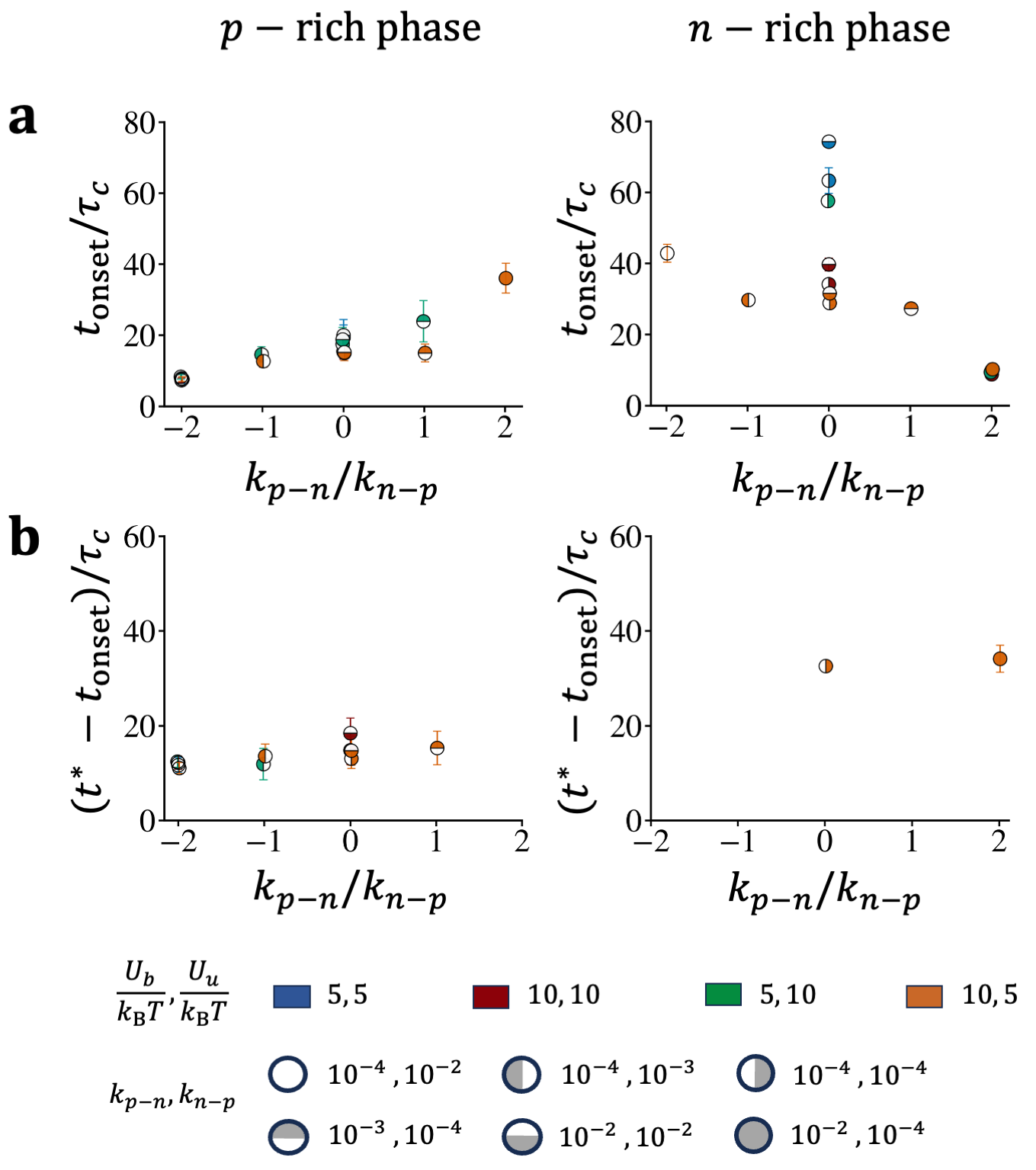}
\caption{\textbf{Onset and duration of phase separation.} The onset of phase separation $t_{\text{onset}}$ (a) and the duration of phase separation $(t^* - t_{\text{onset}})$ (b) scaled by $\tau_c$ (where $\tau_c = (\delta t)_{\text{max}} = 10^{-5}$) are shown with $k_{p-n}/k_{n-p}$. The colors represent different combinations of $U_b$ and $U_u$, and the open, partially filled and filled circles correspond to various cases of $k_{p-n}$ and $k_{n-p}$. The figures in the left and right panels in (a) and (b) correspond to the $p-$rich and $n-$rich phases, respectively. These results correspond to $\alpha = 5$, $\chi_{pn} = 5$, and $\chi_{ps} = \chi_{ns} =  3$. Here $t_{\text{onset}}$ is obtained be averaging the time when volume fraction reaches 0.4 and $t^*$ is similarly obtained when volume fraction reaches 0.6 in the phase separating strips. }  
\label{fig:time_morph_1}
\end{figure}

We report the evolution of characteristic length scales $L_p$ and $L_n$, and the number of stripes $N_p$ and $N_n$ having volume fraction greater than 0.6 to represent the morphology of the $p-$rich and $n-$rich phases, respectively, over time (see \ref{sec:App_c} for how $L_p$, $L_n$, $N_p$ and $N_n$ are obtained). The lower values of $L_p$ and $L_n$ indicate phases forming many narrow stripes with closely spaced domains, whereas their higher values represent phases coarsened into wider domains with larger spacing. Comparisons of different combinations of $U_b$ and $U_u$ are shown in Fig.~\ref{fig:morph}b,c,  \ref{fig:supp_morph_1}, \ref{fig:supp_morph_2}. When $U_b > U_u$ (Fig.~\ref{fig:morph}b and Fig.~\ref{fig:time_morph_1}), the binding time scale is higher whereas unbinding is faster. New cross-link formation events occur less readily than unbinding of the existing cross-links. This weakens the elastic memory of the network phase and allows the domain coarsening of both the $p-$rich and $n-$rich phases with less resistance. The $p-$rich phase coarsens quickly, as reflected in the growth of $L_p$ and rapid saturation of its number of domains, except for a higher value of $k_{p-n}/k_{n-p}$ as discussed earlier (Fig.~\ref{fig:morph}b). The $n-$rich phase also coarsens relatively faster as $L_n$ grows with the merge of finer strips. Although $N_n$ is low or close to zero, as $\phi_n$ does not reach the cut-off value 0.6 within the simulation time window that we could perform, and therefore, the former does not indicate that the $n-$rich regions do not phase separate. The early stage and accelerated coarsening of the $n-$rich phase are observed for a higher value of $k_{p-n}/k_{n-p}$.

In contrast, for the case $U_b < U_u$ (Fig.~\ref{fig:morph}c and Fig.~\ref{fig:time_morph_1}), binding events occur at a faster rate, while unbinding events are suppressed. It is apparent that networks are long-lived after successful formation of more cross-links. Long-lived cross-links mechanically stabilize the morphology and arrest or significantly delay coarsening of both the $p-$rich and $n-$rich phases. This results in formation of finer domains that do not easily merge and stay for longer periods of time. It further indicates slower coarsening as reflected by the slow growth rate of both $L_p$ and $L_n$. Some cases are examined for $U_b = U_u$ (see Fig.~\ref{fig:time_morph_1} and Fig.~\ref{fig:supp_morph_1}). Coarsening is relatively easier for both phases for a higher value of $U_b = U_u$, as progressive build-up of elastic memory from new binding events occurs less rapidly. However phase separation time scale increase for both the $p-$rich and $n-$rich phases for a lower value of $U_b = U_u$. For a lower value of $U_b = U_u$, network remains dynamic, but at the same time it can add up elastic memory rapidly through quick binding events. When the cross-linking parameter $\alpha$, which indirectly represents maximum number of cross-links that can exist for a value of $\phi_n$, is increased, the morphology and phase separation dynamics remain  less affected, except delay is phase separation and coarsening (see Figs.~\ref{fig:supp_morph_2} and \ref{fig:time_morph_2}). The findings indicate that the unbinding kinetics play a stronger role in domain coarsening by suppressing persistence of finer structures when the unbinding time scale is small. However, the binding kinetics govern the stability of the domains by maintaining finer structure or long-time consuming  coarsening when binding time scale is smaller, leading to rapid build up of elastic memory and dynamical arrest. In brief, these findings conclude that the competition between cross-link binding and unbinding kinetics strongly regulate domain sizes and controls the characteristic timescales of coarsening dynamics for both liquid-like and network-like phases.

\section{Discussion}
This work showed a mechanistic route for the onset of condensate aging through a coupled transition in morphology and mechanics. Our model demonstrated that both the morphology  transition and transition in material properties can emerge from the same underlying molecular process: conformational switching followed by dynamic cross-linking. Such modified molecular interactions affect molecular interaction lifetimes which is one of the key factors that can regulate condensate features, as demonstrated by a recent work \citep{debnath2026multi}.  Our modeling assumptions are consistent with previously reported experimental findings \citep{guillen2020rna}, which showed that G3BP molecules go through conformational rearrangement in presence of RNA molecules and form G3BP-RNA stress granules via heterotypic multivalent interactions and cross-linking. Adopting this principle, our minimal continuum model showed how biochemical regulation of molecular switching and cross-linking interactions can translate into various regimes such as rapid early coarsening, slow late-stage coarsening, and  dynamical arrest. 

An implication of this work is that morphology transition can serve as a readily observable signature of pathological states. In a liquid-like state, condensates continue to coarsen and dissolve under different cellular stress conditions. But, in a diseased state, when elastic memory progressively builds up in condensates, coarsening can be significantly suppressed for both liquid-like and network-like phases with persistent arrested structure, shapes and altered rheological properties. Therefore, morphology transition can play role in biomarkers when microrheological measurements are beyond scope and where time-resolved imaging of domain size, number, and onset and duration of maturation can provide indirect evidence of a pathological state.

The former interpretation, however, cannot be generalized  that liquid-like to solid-like transition is an indication of disease progression. Such transitions can also be linked to a healthy process like embryo development \citep{bose2022liquid}. An important distinction is whether such transitions in condensate features are controllable and reversible, or persistent structures suppress remodeling scopes. Our findings therefore suggest that condensate features and their functions are not only governed by phase separation, but also factors affecting molecular interaction lifetimes after condensates form. In the above context, regulating cross-linking interactions among biomolecules appears as a physical strategy that cells prefer to tune condensate fluidity and their functional ability. Hence developing therapeutic strategies to suppress phase separation or shatter condensate domains may not be required. Instead, restoring cross-link turnover and fluidity of condensates may be sufficient to recover their functions. Targeting to modulate interaction lifetimes of condensate forming biomolecules  may therefore offer a plasuible route to reverse pathological arrest, while avoiding to disrupt their physiological activities.

\section*{Data availability}

All study data and derivation details are included in the article. The codebase is available at \url{https://github.com/BHANJANDEBNATH/LLPS_Projects.git}.

\section*{Acknowledgments}

\noindent
The author thanks Profs. V. Kumaran, Parag Katira, Balaji Iyer and Lopamudra Giri for valuable discussions. The author acknowledges the support of PARAM SEVA--IIT Hyderabad under National Supercomputing Mission (NSM) and JICA cluster 
for providing computational support to generate the simulation data. B.D. acknowledges the Department of Science and Technology (DST), Ministry of Science and Technology, India
INSPIRE Faculty award DST/INSPIRE/04/2024/000725. The author also acknowledges Codex Plus 5.5 support to  review, extend and organize simulation and analyses codebase to facilitate parallel processing of ranges of parameters.

\vspace{0.2cm}
\noindent
\textbf{Conflict of interest}: There are no conflicts to declare.

\appendix
\renewcommand{\thesection}{Appendix \Alph{section}}
\setcounter{figure}{0}
\renewcommand{\thefigure}{App.\arabic{figure}}
%\captionsetup[figure]{name=Appendix Figure}
\setcounter{equation}{0}
\renewcommand{\theequation}{A.\arabic{equation}}

\section{Model for $k_{\text{off}}$ in aging condensates}
\label{sec:App_a} 

The transition from liquid-like condensate to aging condensate is not a single mode pathway. Our model hypothesizes that protein molecules that form liquid-like phases can form network-like phases when the biochemical environment changes. The strength and type of interactions among protein molecules alter upon the binding/unbinding interaction of a protein molecule with another molecule in the local environment.  Now there can be two possibilities for further transition. In one case, network phases remain transient due to dynamic cross-linking, implying non-aging. In the second case, condensates age as time elapses and the mechanical response of the network changes with time. Aging of a condensate environment may be due to many reasons, including continuous conformational change of proteins, increased local ordering, network densification via stabilization of cross-links, solvent expulsion, and insufficiency of molecules like molecular chaperones and ATP molecules inside the condensate environment.

We hypothesize that all these former factors that contribute to condensate aging directly or indirectly stabilize the cross-links of the network progressively over time. To model aging and stabilization of cross-links, there may be two modeling approaches. In one approach, a cross-link matures independently according to the time elapsed since its formation. The unbinding rate of a cross-link at time $t$ that formed at time $t^\prime$ varies with its age $t - t^\prime$ in the following phenomenological manner:
\begin{gather}
k_{\text{off}}(x,t;t^\prime) \, = \, k_u \, k_\lambda \, \Big( 1 \, + \frac{t-t^\prime}{\tau_a}\Big)^{-a},
\label{eq:k_off_bond_aging}\\
    k_u \, = \, k_{\text{off}}^0 \, \exp\Big(-\frac{U_u}{\kb T}\Big) \; \; \;  {\rm{and}} \; \; \;  k_\lambda \, = \, \exp \, \Bigg(
    \frac{1}{2} \, \Bigg|\frac{\lambda^2(x,t)}{\lambda^2(x,t^\prime)}
     - 1 \Bigg|\Bigg),
     \label{eq:ku_kl}
\end{gather}
where $\tau_a$ is the characteristic time scale of aging and $a > 0$ governs the strength of aging. The form (\ref{eq:k_off_bond_aging}) implies that the age clock continuously resets with every newly formed cross-link after breakage and rebinding of cross-links, irrespective of the age of the surrounding network. 
It is intuitive that the network, consequently, contains predominantly young cross-links that retain negligible memory of the global waiting time. The assumption that a newly formed cross-link in an aged network has the same initial stability as a cross-link in a fresh network is not physically sound. Therefore, the form (\ref{eq:k_off_bond_aging}) is not pursued here further. 

Rather, we assume aging to be a feature of the whole network rather than an internal clock that resets at every binding event. In a phenomenological manner, we propose the modified unbinding rate of cross-links of a network at time $t$ that is aged for a time $t_s$ as
\begin{equation}
    k_{\mathrm{off}}(x,t_s, t;t^\prime) \, = \, k_u \, 
    k_\lambda, \, \Big( 1 \, + \, \frac{t \, - \, t_s}{\tau_a} \Big)^{-a}. 
    \label{eq:unbinding_rate_aging}
\end{equation}
This modified form of $k_{\text{off}}$ (\ref{eq:unbinding_rate_aging}) can capture the progressive stabilization of long-lived cross-links, indicating that old cross-links are less likely to break than newly formed cross-links. Note that (\ref{eq:unbinding_rate_aging}) incorporates the effect of global aging on the stability of newly formed cross-links in an aging network. The scenario in (\ref{eq:unbinding_rate_aging}) is completely different from that in (\ref{eq:k_off_bond_aging}). In the limit of negligible aging $(t - t_s) \ll \tau_a$, (\ref{eq:unbinding_rate_aging}) reduces to (\ref{eq:unbinding_rate}).

Now, in an aging network, if the network waits at rest for a period $[t = -t_w, t = 0]$ before it experiences any mechanical stress at $t = 0$,  at time $t$, the correct age of the existing cross-links is $(t_w + t)$. Therefore, from (\ref{eq:unbinding_rate_aging}), $k_{\text{off}}(x,t_w,t;t^\prime)$ is
\begin{gather}
k_{\text{off}}(x,t_w, t;t^\prime) \, = \, k_u \, k_\lambda \, \Big( 1 \, + \frac{t_w + t}{\tau_a}\Big)^{-a}.
\label{eq:modified_k_off}
\end{gather}
For $a = 0$, the system  exhibits non-aging behavior and a network remains fluid-like due to dynamic cross-linking, whereas the system ages for $a > 0$.  The dimensionless number $k_u \, \tau_a$ governs the relative aging rate.

\section{Network mechanics}

\subsection{Uniaxial step-strain and stress relaxation}
\label{sec:App_b1} 

\begin{figure}
    \centering
    \includegraphics[width=0.8\linewidth]{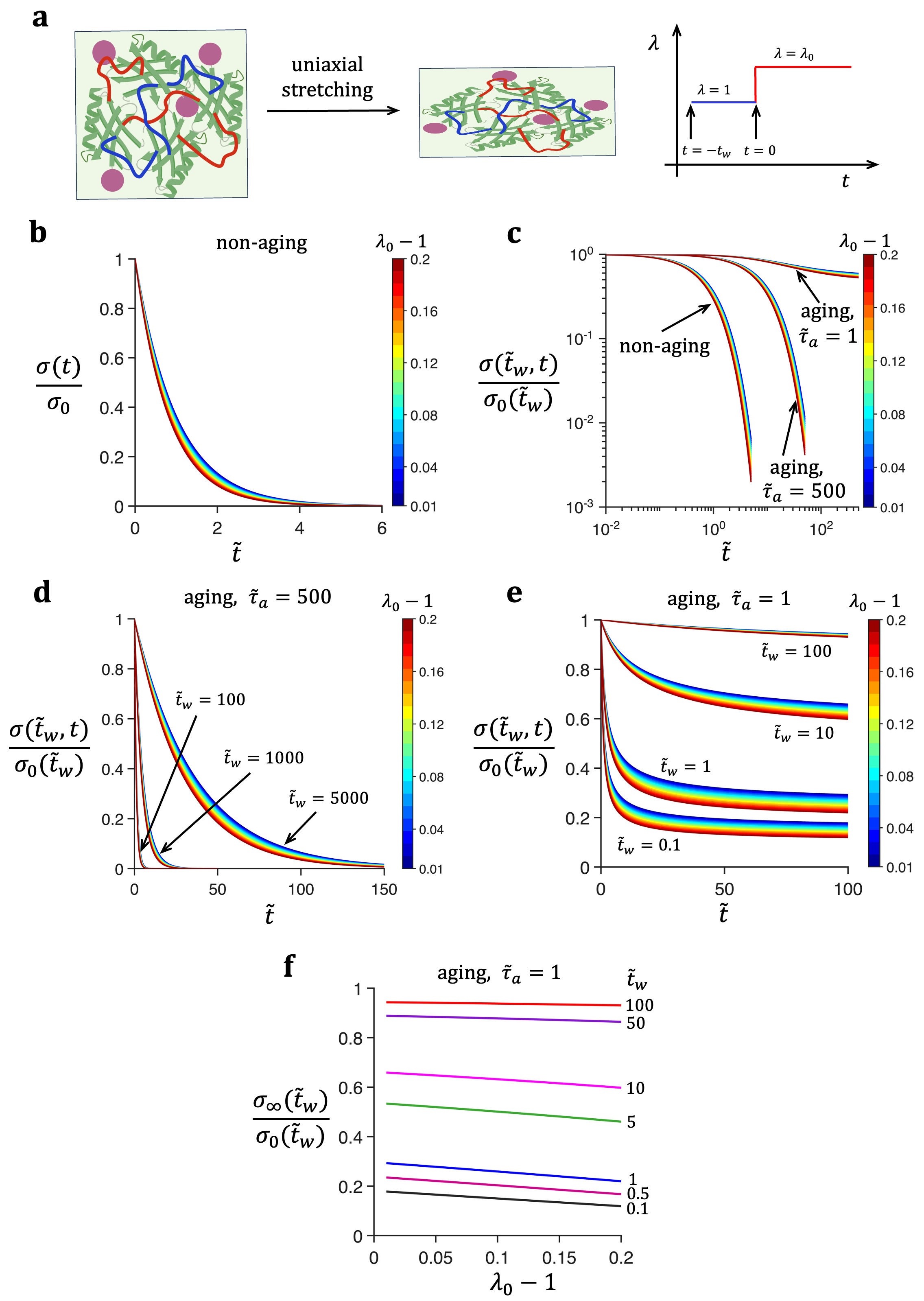}
    \caption{\textbf{Uniaxial step-strain and stress relaxation.} (a) Schematic of step-strain experiment. (b) Exponential decay of the scaled stress $(\sigma(t)/\sigma_0)$ with the scaled time $\tilde{t} = k_u t$ and stress-relaxation of a non-aging network. (c) In log-log scale, comparison of stress relaxation trends from non-aging and aging cases. In (c), for $\tilde{\tau}_a = k_u \tau_a = 500$,  the scaled waiting time $\tilde{t}_w = k_u t_w = 2000$, and for $\tilde{\tau}_a = 1$,  $\tilde{t}_w = 10$. (d) Stress relaxation for different values of $\tilde{t}_w$ when $ \tilde{\tau}_a = 500 \gg 1$. (e) Stress relaxation for different values of $\tilde{t}_w$ and incomplete recovery of an aging network when $\tilde{\tau}_a = 1$. (f) Variation of the scaled residual stress $\sigma_\infty(\tilde{t}_w)/\sigma_0(\tilde{t}_w)$ with the extent of stretch $(\lambda_0 -1)$ for different values of $\tilde{t}_w$ when $\tilde{\tau}_a = 1$. For all results, $\phi_n^0 =1$. }
    \label{fig:stress_relax}
\end{figure}

Consider a network in a stationary state at $t < 0$. A homogeneous uniaxial  instantaneous stretch at $t \geq 0$ is applied to the network (Fig.~\ref{fig:stress_relax}a):
\begin{equation}
    \lambda(t)=
\begin{cases}
1, & -t_w < t<0,\\
\lambda_0 \, = \, 1 + \epsilon_0, & t\ge 0 .
\end{cases}
\label{eq:lambda_strain1}
\end{equation}
When the stretch $\lambda_0$ is held for a duration, we analyze how stress relaxes with time in that duration for non-aging and aging networks.

In a stationary non-aging network, the network remains in equilibrium, and the number of bound cross-links before the network experiences step strain is (see (\ref{eq:nb0}))   
\begin{equation}
    N_b^*|_{t = -t_w} \, = \, N_b^*|_{t = 0}  \equiv \, N_b^* \, = \, \alpha  \, \phi_n^0 \, \frac{k_{\text{on}}}{k_u \, + \, k_{\text{on}}},
    \label{eq:nb_os1_nonag}
\end{equation}
where $k_u = k^0_{\text{off}} \, {\rm exp} \, (-U_u/(k_{\text{B}}T))$ and $k_{\text{on}} = k_b = k^0_b \, {\rm exp} \, (-U_b/(k_{\text{B}}T))$. As $\lambda_0$ remains constant, $\lambda(t) = \lambda(t^\prime) = \lambda_0$ for $t \geq 0$. After instantaneously applying the stretch, the newly formed cross-links at time $ t^\prime$ remain stress-free in the deformed state. Therefore, although $B(t;t^\prime) \neq 0$,  the contribution to the stress due to cross-links that form after application of step strain vanishes due to $
\frac{\lambda_0^2(t)}{\lambda_0^2(t^\prime)} - \frac{\lambda_0(t^\prime)}{\lambda_0(t)}=0
$ (see \ref{eq:stress_xx}). However, the contribution to stress from cross-links at time $t = 0$ just before the stretch is non-zero as they experience a stretch $\lambda_0$. Therefore, from (\ref{eq:stress_xx}), the stress under a step strain would be 
\begin{equation}
    \sigma_{\text{el,tr}}|_{xx}(t;0) \, \equiv \, \sigma(t;0) \, = \,  \, G^* \, A(t;0) \, \Big[\lambda^2_0 \, - \, \frac{1}{\lambda_0} \Big],
    \label{eq:stress_strain1}
\end{equation}
where $G^* \sim (k_{\text B}T/v) \, N_b^*$, and  
\begin{equation}
     \frac{\sigma(t;0)}{\sigma_0}\Bigg|_{\text{non-aging}} \, = \,   A(t;0) ,
    \label{eq:stress_strain2}
\end{equation}
where $\sigma_0 \, = \, G^* \, (\lambda^2_0 \, - \, 1/\lambda_0) $ is the instantaneous stress. From (\ref{eq:unbinding_rate}), (\ref{eq:modified_k_off}), (\ref{eq:ku_kl}) and (\ref{eq:A_B2}), $k_{\text{off}}$ and $A$ for non-aging case would be 
\begin{gather} k_{\text{off}}\Big|_{\text{non-aging}} \, = \, k_u \, k_{\lambda_0} \, , \, \, \, A(t;0)\Big|_{\text{non-aging}} \, = \, \exp \, (- \, k_u \, k_{\lambda_0} \, t) \, ,
    \label{eq:non_aging_k_A} \\
     k_{\lambda_0} \, = \, \exp\Big(\frac{1}{2} \, |\lambda_0^2 \, - \, 1| \Big).
 \label{eq:kl_0}
\end{gather}
The factor (\ref{eq:kl_0}) is due to cross-links present in the network immediately before the step strain. The relative stretch those cross-links experience  after the step strain is $\lambda_0$, which is not valid for newly formed cross-links that form after the stretch at time $t^\prime > 0$. The newly formed cross-links remain force free and therefore, for newly formed cross-links, $k_{\lambda_0} = 1$.

Next we consider an aging network. We assume that the network is in equilibrium at $t = -t_w$ before aging begins. The number of bound cross-links at  time $t = -t_w$ is (see (\ref{eq:nb0})) 
\begin{equation}
    N_b^*\Big|_{t = -t_w} \, \equiv \, N_b^* \, = \, \alpha  \, \phi_n^0 \, \frac{k_{\text{on}}}{k_u \, + \, k_{\text{on}}}.
    \label{eq:nb_os1}
\end{equation}
Now, the network undergoes aging and maturation (in the absence of external force) for a period $t_w$. The equations that govern the evolution of the network  during the period $t_w$ are the following.  At time $t = 0$, the number of bound cross-links would be  (from (\ref{eq:lambda}), (\ref{eq:ntot}), (\ref{eq:crosslink_bal_sol}), (\ref{eq:A_B1}) and (\ref{eq:A_B2}), using (\ref{eq:lambda_strain1}) and setting $\phi_n^0 = 1$), 
\begin{equation}
    \frac{N_b|_{t = 0}}{N_b^*} \, \equiv \, \mathcal{N}(t_w) \ = \, \Bigg[ A(t_w,0) \, + \, \int_{-t_w}^0 \, \md t^\prime \, B(t_w,t^\prime) \Bigg],
    \label{eq:nb_wait_t0}
\end{equation}
where
\begin{gather}
    A(t_w, 0) \,  = \, \Bigg[{\rm exp} \Big(-\int_{-t_w}^0 \, \md s \, k_{\text{off}}(t_w,s) \Big)\Bigg], 
    \label{eq:A_B1_aging1}\\ 
    B(t_w, t^\prime) \,  = \, 
    k_{\text{on}} \, \frac{\Big(\alpha \,  \, - \, N_b(t^\prime)\Big)}{N_b^*} \, \, \Bigg[{\rm exp} \Big(-\int_{t^\prime}^0 \, \md s \, k_{\text{off}}(t_w,s) \Big)\Bigg],
    \label{eq:A_B2_aging1}\\
    k_{\mathrm{off}}(t_w,s) \, = \, k_u \, \Big( 1 \, + \, \frac{t_w \, + \, s}{\tau_a} \Big)^{-a}.
    \label{eq:koff_tw_st}
\end{gather}
After applying step strain $\lambda(t) = \lambda_0$, the stretch is kept constant for a period after $t > 0$. Following (\ref{eq:stress_strain1}), in a similar manner, we can write the expression for stress $\sigma(t_w,t)$ for an aging network under step strain:
\begin{equation}
    \sigma(t_w, t;0) \, = \,  \, G^* \, \mathcal{N}(t_w)\, A(t_w,t;0) \, \Big[\lambda^2_0 \, - \, \frac{1}{\lambda_0} \Big].
    \label{eq:stress_strain1_ag}
\end{equation}
Therefore, 
\begin{gather}
     \frac{\sigma(t_w,t;0)}{\sigma_0(t_w)}\Bigg|_{\text{aging}} \, = \,   A(t_w, t;0) , \, \, \, {\rm where} \, \, \, \sigma_0(t_w) \, = \, G^* \, \mathcal{N}(t_w)\, \Big[\lambda^2_0 \, - \, \frac{1}{\lambda_0} \Big],
    \label{eq:stress_strain2_ag}
\end{gather}
where
\begin{gather}
    A(t_w, t;0) \,  = \, \Bigg[{\rm exp} \Big(-\int_{0}^t \, \md s \, k_{\text{off}}(t_w,s) \Big)\Bigg], 
    \label{eq:A_B1_aging22}\\ 
    k_{\mathrm{off}}(t_w,s) \, = \, k_u \, k_{\lambda_0} \Big( 1 \, + \, \frac{t_w \, + \, s}{\tau_a} \Big)^{-a}.
    \label{eq:koff_tw_st_2}
\end{gather}
From (\ref{eq:A_B1_aging22}) and (\ref{eq:koff_tw_st_2}), we get
\begin{gather}
    A(t_w, t;0)\Big|_{\text{aging}} \, = \, \exp\Bigg(- \, k_u \, k_{\lambda_0} \, \frac{\tau_a}{1-a} \, \Bigg\{ \Big(1 + \frac{t_w + t}{\tau_a} \Big)^{1-a} \, - \, \Big(1 + \frac{t_w}{\tau_a} \Big)^{1-a} \Bigg\}\Bigg).
    \label{eq:aging_k_A}
\end{gather}
From (\ref{eq:stress_strain2_ag}),  the residual stress would be 
\begin{equation}
    \sigma(t_w,t \rightarrow\infty) \, \equiv \, \sigma_\infty(t_w) \,  = \, \sigma_0(t_w) \, \exp\, \Bigg[ \, k_u \, k_{\lambda_0} \, \frac{\tau_a}{1 - a} \, \Big(1 + \frac{t_w}{\tau_a}\Big)^{1-a}\Bigg]. 
    \label{eq:res_stress_corr}
\end{equation}
We scale the variables having time dimension as follows:
\begin{gather}
    \tilde{t} = k_u t \, ; \, \, \, \,  \tilde{\tau}_a = k_u \tau_a \, ; \, \, \, \tilde{t}_w = k_u t_w.  \label{eq:time_scaling}
\end{gather}
Using the former scaling, (\ref{eq:stress_strain2_ag}) and (\ref{eq:res_stress_corr}) reduce to
\begin{gather}
     \frac{\sigma(t_w,t)}{\sigma_0(t_w)} \Bigg|_{\text{aging}}\, = \, \exp\Bigg(- \, k_{\lambda_0} \, \frac{\tilde{\tau}_a}{1-a} \, \Bigg\{ \Big(1 + \frac{\tilde{t}_w + \tilde{t}}{\tilde{\tau}_a} \Big)^{1-a} \, - \, \Big(1 + \frac{\tilde{t}_w}{\tilde{\tau}_a} \Big)^{1-a} \Bigg\}\Bigg), \label{eq:stress_relax_aging_scaled}\\  
     \sigma_\infty(t_w) \,  = \, \sigma_0(t_w) \, \exp\, \Bigg[ \, k_{\lambda_0} \, \frac{\tilde{\tau}_a}{1 - a} \, \Big(1 + \frac{\tilde{t}_w}{\tilde{\tau}_a}\Big)^{1-a}\Bigg],    \label{eq:res_stress_corr_scaled}
\end{gather}
Here (\ref{eq:res_stress_corr_scaled}) reveals the following insights. The residual stress $\sigma_\infty$ depends on $t_w$, constrained to whether $\tilde{\tau}_a = k_u \tau_a \gg 1$ or $\tilde{\tau}_a = k_u \tau_a \sim 1$. Although a system is aging, (\ref{eq:res_stress_corr_scaled}) indicates that the residual stress  $\sigma_\infty \rightarrow 0$ for $a > 1$ and $\tilde{\tau}_a \gg 1$, unless the scaled waiting time is significantly longer than the scaled characteristic time scale for aging, i.e., $\tilde{t}_w \gg \tilde{\tau}_a$. However, the residual stress nevertheless remains nonzero for finite parameter values in (\ref{eq:res_stress_corr_scaled}). If $\tilde{\tau}_a \sim 1$, $\sigma_\infty > 0$ for $a > 1$, irrespective of $\tilde{t}_w > 0$. Therefore, for the same condition $a > 1$, there can be two regimes: 

(1) $\tilde{\tau}_a \gg 1$: stress can relax and recovery can occur over a longer period of time, like a Maxwell fluid, but the relaxation time strongly varies with the waiting time $t_w$. This behavior is similar to that of the aging visco-elastic Maxwell like fluid. 

(2) $\tilde{\tau}_a \sim 1$ stress may relax, but with  incomplete recovery and non-zero residual stress as a function of the waiting time $t_w$. This behavior is close to that of the characteristic of a visco-elastic solid. 

These former insights encourage us to choose $a > 1$, and thus we set $a = 1.5$ in this study. However, more rigorous study and a suitable model for $a$ would be  required to connect it with the internal structural changes of the network during maturation or aging. This topic merits more attention and is a part of future work. 

From (\ref{eq:stress_strain2}) and (\ref{eq:stress_relax_aging_scaled}), it is evident that non-aging networks show exponential decay, whereas aging networks can show non-exponential decay governed by a time-dependent unbinding rate (see Fig.~\ref{fig:stress_relax}b,c). Note that (\ref{eq:stress_strain2}) and (\ref{eq:non_aging_k_A}) infer $\sigma(t\rightarrow\infty)/\sigma_0 \, \rightarrow \, 0$, indicating quick and complete recovery, as reflected in Fig.~\ref{fig:stress_relax}b. The stress relaxation trends in Fig.~\ref{fig:stress_relax}b infer that the non-aging network behaves as a visco-elastic Maxwell like fluid.
An aging system also reflects complete, however, delayed recovery, when $\tilde{\tau}_a = k_u \tau_a \gg 1$ (Fig.~\ref{fig:stress_relax}c). The rate of aging is slower for the case $ \tilde{\tau}_a = 500$, and stress relaxation occurs over longer periods of time  with an increase in the scaled waiting time $\tilde{t}_w = k_u t_w$ (Fig.~\ref{fig:stress_relax}d). The stress relaxation trends at different $\tilde{t}_w = k_u t_w$ in Fig.~\ref{fig:stress_relax}d implies that the relaxation time depends on the waiting time $t_w$, implying an aging visco-elastic Maxwell like fluid.     

When $\tilde{\tau}_a = k_u \tau_a = 1$, the aging time scale $\tau_a$ is set to the same value as the cross-link unbinding time scale $1/k_u$ in a fresh network, implying an increase in the speed of aging, compared to $\tilde{\tau}_a \gg 1$ (Fig.~\ref{fig:stress_relax}c). Aging progresses very rapidly with a small increase in  $\tilde{t}_w$ (Fig.~\ref{fig:stress_relax}e), consequently an increase in scaled residual stress (Fig.~\ref{fig:stress_relax}f). When $\tilde{\tau}_a \sim 1$, the higher value of the scaled residual stress  close to 1 (Fig.~\ref{fig:stress_relax}f) at a higher value of  $\tilde{t}_w$  indicates that the non-dissipating stress is more like elastic when the strain is held constant. We speculate that the aging network behaves as a visco-elastic solid-like material when it ages significantly and very rapidly. We next use oscillatory-strain test and  active microrheology-based creep test to determine whether the model captures the transition from a relaxing transient network, through an aging viscoelastic-fluid, to a dynamically arrested viscoelastic-solid.

\subsection{Oscillatory strain: storage and loss moduli}
\label{sec:App_b2} 

We employ the following experimental protocol for an oscillatory strain at $t \geq 0$ %(see Fig.~\ref{fig:os_strain}a):
\begin{equation}
    \lambda(t)=
\begin{cases}
1, & -t_w < t < 0,\\
1 \, + \, \epsilon_0 \, \sin \omega t, & t\ge 0 .
\end{cases}
\label{eq:oscillatory_strain}
\end{equation} 
From (\ref{eq:stress_xx}) and (\ref{eq:nb_wait_t0}), the stress expressions for non-aging and aging networks under oscillatory strain would be, respectively, 
\begin{gather}
     \frac{\sigma(t)}{G^*} \, =    \,  A(t;0) \,  \Bigg[\lambda^2(t)
    \, - \, \frac{1}{\lambda(t)}\Bigg] \,
     +  \, \int_0^t \, \md t^\prime \, B(t;t^\prime)  \, \Bigg[ 
    \frac{\lambda^2(t)}{\lambda^2(t^\prime)} \, - \, \frac{\lambda(t^\prime)}{\lambda(t)} 
    \Bigg] ,
     \label{eq:stress_nonag_os}\\
     \frac{\sigma(t_w,t)}{G^*} =    \, \mathcal{N}(t_w) \, \Bigg( A_w(t_w,t;0) \,  \Bigg[\lambda^2(t)
    \, - \, \frac{1}{\lambda(t)}\Bigg] \,
     +  \, \int_0^t \, \md t^\prime \, B_w(t_w,t;t^\prime)  \, \Bigg[ 
    \frac{\lambda^2(t)}{\lambda^2(t^\prime)} \, - \, \frac{\lambda(t^\prime)}{\lambda(t)} 
    \Bigg] \Bigg),
     \label{eq:stress_xx_ag}
\end{gather}
where $G^* \sim (k_{\text B}T/v) \, N_b^*$, and 
\begin{gather}
    A_w(t_w,t;0) \,  = \, \Bigg[{\rm exp} \Big(-\int_{0}^t \, \md s \, k_{\text{off}}(t_w,s;0) \Big)\Bigg], 
    \label{eq:A_B1_aging2}\\ 
    B_w(t_w, t; t^\prime) \,  = \, 
    k_{\text{on}} \, \frac{\Big(\alpha \, \phi_n (t^\prime)   \, - \, N_b(t^\prime)\Big)}{N_b^* \, \mathcal{N}(t_w)} \, \, \Bigg[{\rm exp} \Big(-\int_{t^\prime}^t \, \md s \, k_{\text{off}}(t_w,s;t^\prime) \Big)\Bigg],
    \label{eq:A_B2_aging2}\\
    k_{\mathrm{off}}(t_w,s;t') \, = \, k_u \, \exp \, \Bigg(
    \frac{1}{2} \, \Bigg|\frac{\lambda^2(s)}{\lambda^2(t^\prime)}
     - 1 \Bigg|\Bigg) \, \Big( 1 \, + \, \frac{t_w \, + \, s}{\tau_a} \Big)^{-a}.
    \label{eq:koff_tw}
\end{gather}

We use the form in  (\ref{eq:stress_nonag_os}) and (\ref{eq:stress_xx_ag})--(\ref{eq:koff_tw}) to obtain the storage modulus $G^\prime$ and the loss modulus $G^{\prime \prime}$ of the non-aging and aging networks that age for different waiting times $t_w$. For oscillatory strain analysis, we choose small amplitude $\epsilon_0 = 0.01$, so that linear viscoelasticity theory can be applied to extract  $G^\prime$ and $G^{\prime \prime}$. For the case of a non-aging network, we integrate (\ref{eq:stress_nonag_os}) from $t = 0$ to $N_c$  cycles 
\begin{gather}
G^\prime \, = \,  \frac{1}{N_c \, \pi \, \epsilon_0} \, 
\int_0^{(2\pi/\omega) \, N_c}
\sigma(t) \, \sin(\omega t) \, \omega \, \md t,
\label{eq:Gp_nonag}\\
G^{\prime \prime}
 \, = \, 
\frac{1}{N_c \, \pi \, \epsilon_0}
\int_0^{(2\pi/\omega)N_c}
\sigma(t)\cos(\omega t)\,\omega\,\md t.
\label{eq:Gdp_nonag}
\end{gather}
After nondimensionalization using (\ref{eq:time_scaling}),  the integrals in (\ref{eq:Gp_nonag}) and (\ref{eq:Gdp_nonag}) are performed using trapezoidal time integration; %with 256 time points per cycle and $\delta\tilde{t}\leq 0.025$;
measurements are performed between $N_c =$ 2 to 5 cycles for each frequency $\omega$.

For the case of an aging network, we follow a different protocol as the stress response changes during the observation and measurement interval.  The instantaneous values of the 
moduli are extracted using analytic signals and using (\ref{eq:stress_xx_ag})--(\ref{eq:koff_tw}) \citep{takaki2023theory}. The analytic stress and strain are defined as
\begin{gather}
\sigma_a(t) \, = \, 
 \, \sigma(t) \, + \, i \, \mathcal{H} \, [\sigma(t)], \, \, \, \,  \epsilon_a(t) \, = \,  \epsilon(t) \, + \, i \, \mathcal{H}[\epsilon(t)],
\label{eq:analytic_stress_strain}
\end{gather}
where $\mathcal{H}$ denotes the Hilbert transform. The instantaneous complex modulus is $G_a^*(t)  = \sigma_a(t)/\epsilon_a(t)$. The instantaneous moduli are obtained by averaging $G_a^*(t)$ over one oscillation period centered around time $t = t_{\text{obs}}$:
\begin{gather}
\overline{G^*_a} \, 
= \, 
\frac{1}{T} \, 
\int_{t_{\mathrm{obs}} - T/2}^{t_{\mathrm{obs}}+T/2}
G^*_a(t) \, \md t,
\qquad
T=\frac{2\pi}{\omega},
\label{eq:stress_analytic_av}\\
G^\prime \, = \, \operatorname{Re}\left[\, \overline{G^*_a} \, \right ],
\qquad
G^{\prime \prime} \, = \,  \operatorname{Im}\left[ \, \overline{G^*_a} \, \right].
\label{eq:gp_gdp_ag}
\end{gather}
A common observation time is chosen for all measurements at different frequencies and parameters chosen; its scaled value is $\tilde{t}_{\text{obs}} = 10 \pi$. In experiments with different $\omega$, the aging envelope, which is the amplitude ($\sigma^*$) and phase difference ($\delta_\sigma$) of the stress response $\sigma \sim \sigma^* \sin (\omega t + \delta_\sigma)$, must vary slowly compared to the varying sinusoidal part of the strain as the material ages. Therefore, there must be a lower bound of frequency  $\omega_l$ to analyze the stress response in the case of aging. We employ the Bedrosian criterion to omit frequencies that  are less than $\omega_l \sim |\frac{\md}{\md t} \, \ln (1 + (t_w + t_{\text{obs}} +  t)/\tau_a)^{-a}|$, where the  factor $(1 + (t_w + t)/\tau_a)^{-a}$ represents the rate of aging (see (\ref{eq:modified_k_off})). To be on safe side, $\omega_l = 10 \, a/(\tau_a + t_w + t_{\text{obs}})$. By setting $t_w = t_{\text{obs}} = 0$ at which the rate of change of the stress envelope is the largest, a waiting-time-independent lower bound would be $\omega_l = 10 \, a/\tau_a$. For example,  when $a = 1.5$ and the scaled value $\tilde{\tau}_a = k_u  \tau_a = 500$, the lower bound  of $\tilde{\omega}_l = \omega_l/k_u$ is obtained as 0.03, and hence, analyses at frequencies below this value are not reported for the corresponding set of parameters.

\subsection{Active microrheology-based creep test}
\label{sec:App_b3}

Consider a bead embedded within a transient network. The movement of an optical trap applies force on the bead to displace, and at the same time the network resists the motion of the bead. %(see Fig.~\ref{fig:ac_creep}a).
Assuming the optical trap as an elastic energy well,  the force balance on the bead is 
\begin{equation}
    \xi_b \, \frac{\md x_b}{\md t} \, = \, k_{\text{tr}} \, (x_{\text{tr}} \, - \, x_b) \, - \, F_{n}(t),
    \label{eq:bead_force_bal}
\end{equation}
where $\xi_b$ is the damping coefficient associated with bead displacement in the network medium, $k_{\text{tr}}$ is the stiffness of the trap, $x_{\text{tr}}$ is the center of the trap, $x_b$ is the position of the bead, and $F_n$ is the resistance force on the bead due to the deformation of the network when the bead moves through it. Before solving for $x_b$ using (\ref{eq:bead_force_bal}), we derive $F_n$ from the virtual work in a phenomenological manner. If the effective volume of deformation due to bead movement is $v_{\text{eff}}$, the stored elastic energy is $\mathcal{E}_n = v_{\text{eff}} \, f_{\text{el,tr}}$, and  we can write $F_n = \frac{\partial \mathcal{E}_n}{\partial x_b}$. The relation between the local stretch $\lambda$ and the bead displacement $x_b$ is $\lambda = 1 + (x_b/l_n)$, where $l_n$ is an effective length scale for deformation. Using the former, the final expression for $F_n$ is
\begin{equation}
    F_n(t) \, = \, c_n \, \frac{\sigma(t_w,t)}{\lambda},
    \label{eq:network_force}
\end{equation}
where $c_n = (v_{\text{eff}}/l_n)$ is a geometric constant associated with bead size and shape. We use (\ref{eq:stress_nonag_os}) and (\ref{eq:stress_xx_ag}) along with (\ref{eq:bead_force_bal}) and (\ref{eq:network_force}) to obtain $x_b(t)$ for non-aging and aging networks. The experiment protocol is the following:
\begin{equation}
    x_{\text{tr}} \, = \, 
\begin{cases}
0, & -t_w < t < 0,\\
v_{\text{tr}} \, t, & 0 \leq t < t_r 
\end{cases}
\label{eq:creep}
\end{equation} 
where $v_{\text{tr}}$ is the trap velocity. After time $t_r$, the trap is released, implying $k_{\text{tr}} = 0$ and (\ref{eq:bead_force_bal}) reduces to 
\begin{equation}
    \xi_b \, \frac{\md x_b}{\md t} \, = \, \, - \, F_{n}(t).
    \label{eq:bead_force_bal_rel}
\end{equation}
Scaling displacements by $l_n$, time by $1/k_u$ and force by $c_n G^*$, (\ref{eq:bead_force_bal}) can be reduced to
\begin{equation}
    \tilde\xi_b \, \frac{\md \tilde x_b}{\md \tilde t} \, = \, \tilde k_{\text{tr}} \, (\tilde x_{\text{tr}} \, - \, \tilde x_b) \, - \, \frac{1}{\lambda} \, \frac{\sigma}{G^*},
    \label{eq:bead_force_bal_scaled}
\end{equation}
where $\tilde\xi_b = (\xi_b l_n k_u)/(c_n G^*)$ and $\tilde k_{\text{tr}} = (k_{\text{tr}} l_n )/(c_n G^*)$. Note that for a large value of $\tilde\xi_b$, the bead would move slowly followed by smooth recoil, whereas the bead would respond quickly followed by fast recoil for a smaller value of $\tilde\xi_b$.

\section{Phase separation and morphology: Methods} 
\label{sec:App_c}

\subsection{Simulation protocol}
\label{sec:App_c1}

The variables, model parameters and equations  (see table~\ref{tab:mod_eqns}) are dimensionless. Note that length, time, molecular volume, energy, and stress are scaled by $\ell_g$, $t_g$, $v$, $E_g$, and $\sigma_g$, respectively. Without loss of generality, we set $\ell_g = 1$, $t_g = (1/\kon^0) = (1/\koff^0) = 1$, $v = 1$,  $E_g = \kb T = 1$, and $\sigma_g = E_g/v = 1$. The diffusivity, interfacial coefficient and rate constants are scaled by $\ell_g^2/t_g$, $\ell_g^2$, and $1/t_g =  \kon^0 = \koff^0$, respectively. The scaled diffusivity values in the $p-$rich and $n-$rich phases are set to $D_p = 0.1$ and $D_n = D_p/2$; similarly, the scaled values of the interfacial coefficient is $\kappa = 10^{-2}$ and the rate of diffusion of cross-links in networks is $k_{\text{diff}} = 5 \times  10^{-2}$. The scaled domain size is chosen as $L = 20$.  

The model equations  (see table~\ref{tab:mod_eqns}) are solved in a one-dimensional periodic domain $x \in [0,L]$ by discretizing the space into equally spaced $N = 1024$ points and using the Fourier pseudo-spectral method \citep{shrinivas2021phase}. Spatial derivatives are evaluated using Fourier transforms. For example, if a function $f(x)$ is expressed as $f(x)=\sum_k \hat{f}_k \, \, e^{ikx}$ where $k$ represents Fourier wavenumbers, then the spatial derivatives are obtained as
\begin{gather}
    \widehat{\partial_x f} = ik \hat{f}_k,
\qquad
\widehat{\partial_{xx}f}=-k^2\hat f_k,
\qquad
\widehat{\partial_{xxxx}f}=k^4\hat f_k.
\end{gather}
We implement a first order time implicit (to handle linear terms) -  explicit (to handle non-linear terms) (IM-EX) method to solve time evolution of $\phi_p$ and $\phi_n$ fields \citep{ascher1997implicit}. The balance equations for $\phi_p$ and $\phi_n$ are discretized as
\begin{gather}
    \frac{\phi_p^i(t+\delta t)-\phi_p^i(t)}{\delta t} \, = \, \mathcal{L}\big[\phi_p^i(t+\delta t) \big] \, + \, \mathcal{N}\big[ \phi_p^i(t) \big], \nonumber \\
    \frac{\phi_n^i(t+\delta t)-\phi_n^i(t)}{\delta t} \, = \, \mathcal{L}\big[\phi_n^i(t+\delta t) \big] \, + \, \mathcal{N}\big[ \phi_n^i (t) \big],
    \label{eq:discret}
\end{gather}
where $i$ represents a discretized node, $\mathcal{L}$ and $\mathcal{N}$ are linear and non-linear operators, respectively. Here, $\mathcal{L}\big[\phi_{(p,n)}\big] \, = \, - \, M_{(p,n)} \, \kappa \, \, \partial_{xxxx}\phi_{(p,n)}$; $\mathcal{N}\big[ \phi_p\big] \, = \, 
\partial_x
\left(
M_p \, (\phi_p)^2 \, \,  \partial_x\mu_p
\right)
\, - \, \mathcal{R} \, 
- \, \mathcal L\big[\phi_p\big]
$; $\mathcal{N}\big[ \phi_n\big] \, = \, 
\partial_x
\left(
M_n \, (\phi_n)^2 \, \,  \partial_x\mu_n
\right)
\, -\, \partial_x (M_n \phi_n \partial_x (\sigma_{xx})) \, + \, \mathcal{R} \, 
- \, \mathcal L\big[\phi_n\big]
$. In Fourier space, (\ref{eq:discret}) can be transformed to 
\begin{gather}
    \widehat{\phi_p^i}(t+\delta t) = \, \frac{\widehat{\phi_p^i}(t) \, + \, \delta t \, \, \widehat{\mathcal{N}_p^i}(t)}{1 \, + \, \delta t \, M_p^i \, \kappa \, k^4} , \nonumber \\
    \widehat{\phi_n^i}(t+\delta t) = \, \frac{\widehat{\phi_n^i}(t) \, + \, \delta t \, \,\widehat{\mathcal{N}_n^i}(t)}{1 \, + \, \delta t \, M_n^i \, \kappa \, k^4}.
    \label{eq:discret_fourier}
\end{gather}
We employ an adaptive time stepping scheme based on the maximum change in $\phi_p$ and $\phi_n$ as follows. The increment is defined as 
\begin{gather}
    \Delta_\phi = \max \left\{ \left\|\phi_p(t+\delta t)-\phi_p(t) \right\|_\infty, \, \left\|\phi_n(t+\delta t)-\phi_n(t) \right\|_\infty \right\}.
    \label{eq:incre_tol}
\end{gather} 
If $\Delta_\phi > 2\varsigma_{tol}$, the current simulation step is rejected and replaced by
\[
\delta t \leftarrow \max\!\left(\frac{\delta t}{2},(\delta t)_{\min}\right).
\]
If $\Delta_\phi < \varsigma_{tol}/2$, it is replaced by
\[
\delta t \leftarrow \min\!\left(1.2\delta t,(\delta t)_{\max}\right).
\]
Here we choose $\varsigma_{tol}=10^{-4}$, $(\delta t)_{\max}=10^{-5}$, and $(\delta t)_{\min}=10^{-8}$. After each accepted simulation time step, the overall balance, $\int_0^L \, (\phi_p + \phi_n) \, 
\md x = $ constant, is checked. The simulations are run for a maximum number of timesteps  $10^7$. We specify the initial compositions as random perturbations about homogeneous reference states,
\begin{gather}
    \phi_p(x,t = 0) = \phi_p^0 (x) = \bar{\phi}_p^0 + 0.01 \, \eta (x) ; \, \, \, \, \,  \phi_n(x,0) = \phi_n^0 (x) =  \bar{\phi}_n^0 + 0.01 \, \eta (x),
    \label{eq:ini}
\end{gather}
where $\bar{\phi}_p^0 = 0.35$ and $\bar{\phi}_n^0 = 0.35$ are the average volume fractions of the $p-$rich and $n-$rich phases, respectively; $\eta$ is a normally distributed random field with zero mean which contributes as random perturbations triggering spinodal decomposition.  Note that the total number of cross-links possible at equilibrium is  $N_b^0(x)
\, = \, \alpha \, \phi_n^0(x) \, 
\frac{k_{\mathrm{on}}}
{k_{\mathrm{off}}(x,0) \, + \, k_{\mathrm{on}}}$. But, to initialize different simulations for different sets of $U_b$ and $U_u$ with the same initial cross-link density, we choose $N_b^0(x)=0.1 \, \alpha \, \phi_n^0(x)$ as the initial  number of cross-links in all simulations. This choice can quantitatively affect the initial elastic modulus of networks and the onset of phase separation.

\subsection{Analysis of simulation data} 
\label{sec:App_c2}

\subsubsection{$L_p$, $L_n$, $N_p$, $N_n$ calculation}
\label{sec:App_c21}

Characteristic domain lengths $L_p$ and $L_n$ are calculated from the structure factor in the following manner. For each time snapshot and for any phase, we first calculate 
\begin{gather}
    \delta \phi(x,t) = \phi(x,t) - \langle \phi(t) \rangle, 
    \label{eq:mean_sub}
\end{gather}
where $\langle \phi(t) \rangle = \int \phi \, \md x$. We then compute the Fourier transform
\begin{gather}
    \tilde{\phi}(q,t) = \mathcal{F}[\delta \phi(x,t)].
    \label{eq:struc}
\end{gather}
The structure factor is defined as $S(q,t) = |\tilde{\phi}(q,t)|^2$. For positive wave vectors, the spectral average wave vector is 
\begin{gather}
    \bar{q}(t) = \frac{\sum_{q>0}q \, S(q,t)}{\sum_{q>0}S(q,t)}.
\end{gather}
We obtain the characteristic lengths using the following
\begin{gather}
    L(t) = \frac{2 \pi}{\bar{q}(t)}.
    \label{eq:domain_length}
\end{gather}
For each time snapshot, we calculate $N_p$ and $N_n$ as the numbers of domains of phase separating $p-$rich and $n-$rich strips, respectively, that reach a cut-off value of the volume fraction $\phi_p = 0.6$ and $\phi_n = 0.6$, respectively, within the overall time window of simulations.  

\subsubsection{$t_{\text{onset}}$ and $t^*$ calculation}
\label{sec:App_c22}

From the one dimensional phase fields of $\phi_p$ and $\phi_n$, phase separating stripes are identified from the full time history in each simulation. The spatial locations and corresponding time points were obtained from phase fields when the values of $\phi_p$ and $\phi_n$ in the phase-separating stripes of $p-$rich and $n-$rich stripes, respectively, cross a value 0.4. The corresponding time values are denoted as $t_{\text{onset}}$. The time $t^*$ was then calculated for the same stripes when a value of 0.6 was reached. As simulations are computationally expensive, the longest duration that the simulations could be run is $t/\tau_c = 80$, where $\tau_c$ is a nominal time scale. In this time window, $\phi_p$ and $\phi_n$ have not reached 1 in many cases; that is why, currently the choice of setting 0.4 and 0.6 is arbitrary. The reported values of $t_{\text{onset}}$ and  $t^*-t_{\text{onset}}$ are the mean of all detected stripes for each simulation, and the error bars correspond to 95\% confidence limits. Note that for cases such as a stripe crossing 0.4 but later not crossing 0.6, are only included in calculation of $t_{\text{onset}}$.

\begin{figure}
\centering
\includegraphics[width=0.8\linewidth]{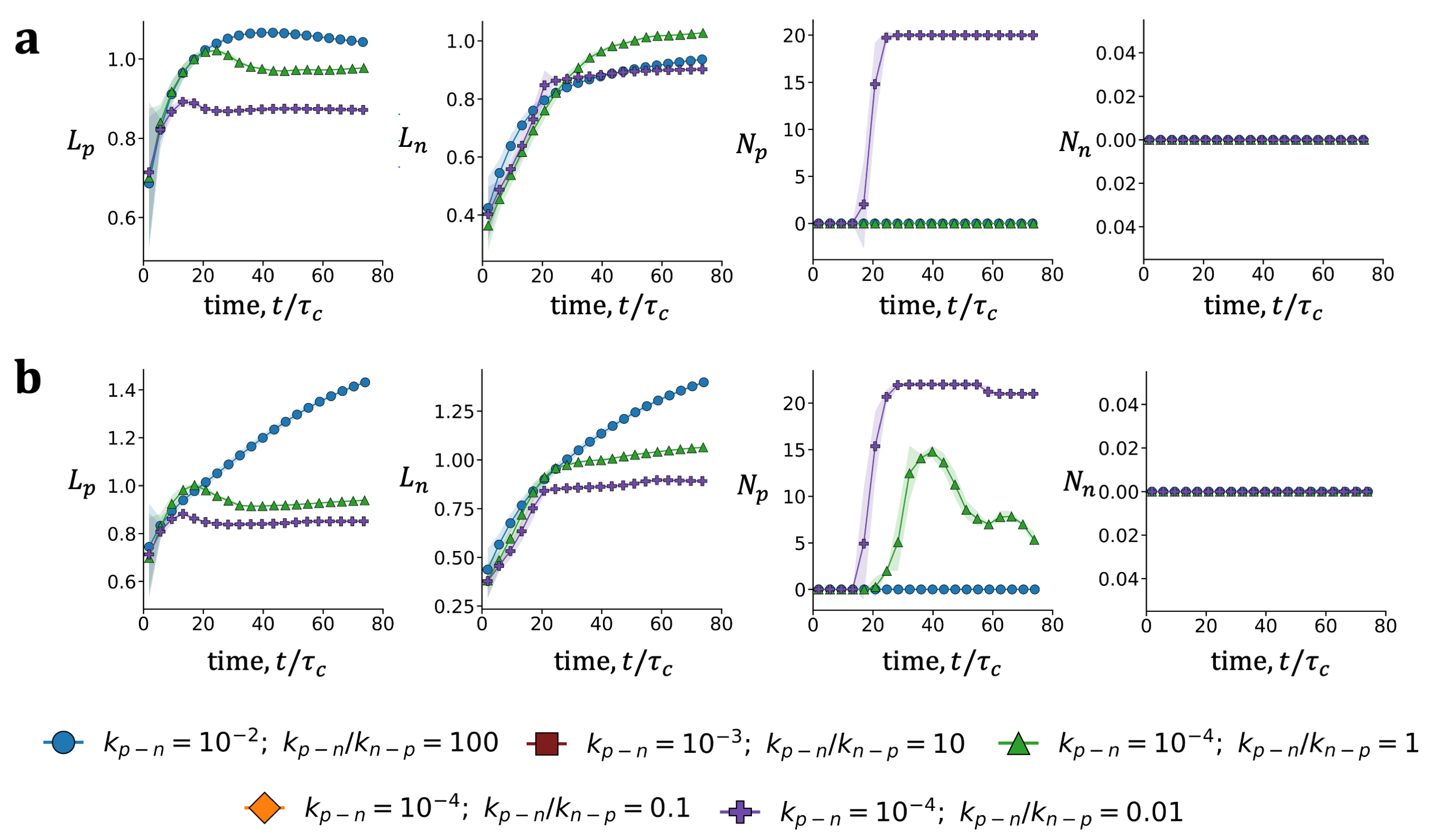}
\caption{\textbf{Morphology.}  In (a) and (b), the characteristic length scales $L_p$ and $L_n$ and number of domains $N_n$ and $N_n$ of the $p-$rich and $n-$rich phases, respectively are plotted over time $t/\tau_c$ (where $\tau_c = (\delta t)_{\text{max}} = 10^{-5}$). In (a) $U_b/\kb T = 5$, $U_u/\kb T = 5$ and in (b), $U_b/\kb T = 10$, $U_u/\kb T = 10$. These results correspond to $\alpha = 5$, $\chi_{pn} = 5$, and $\chi_{ps} = \chi_{ns} =  3$.  In (b) and (c), $N_p$ and $N_n$ represent number of phase-separating strips that reach the volume fraction 0.6 within the time frame the simulations are performed.  }
\label{fig:supp_morph_1}
\end{figure}

\begin{figure}
\centering
\includegraphics[width=0.8\linewidth]{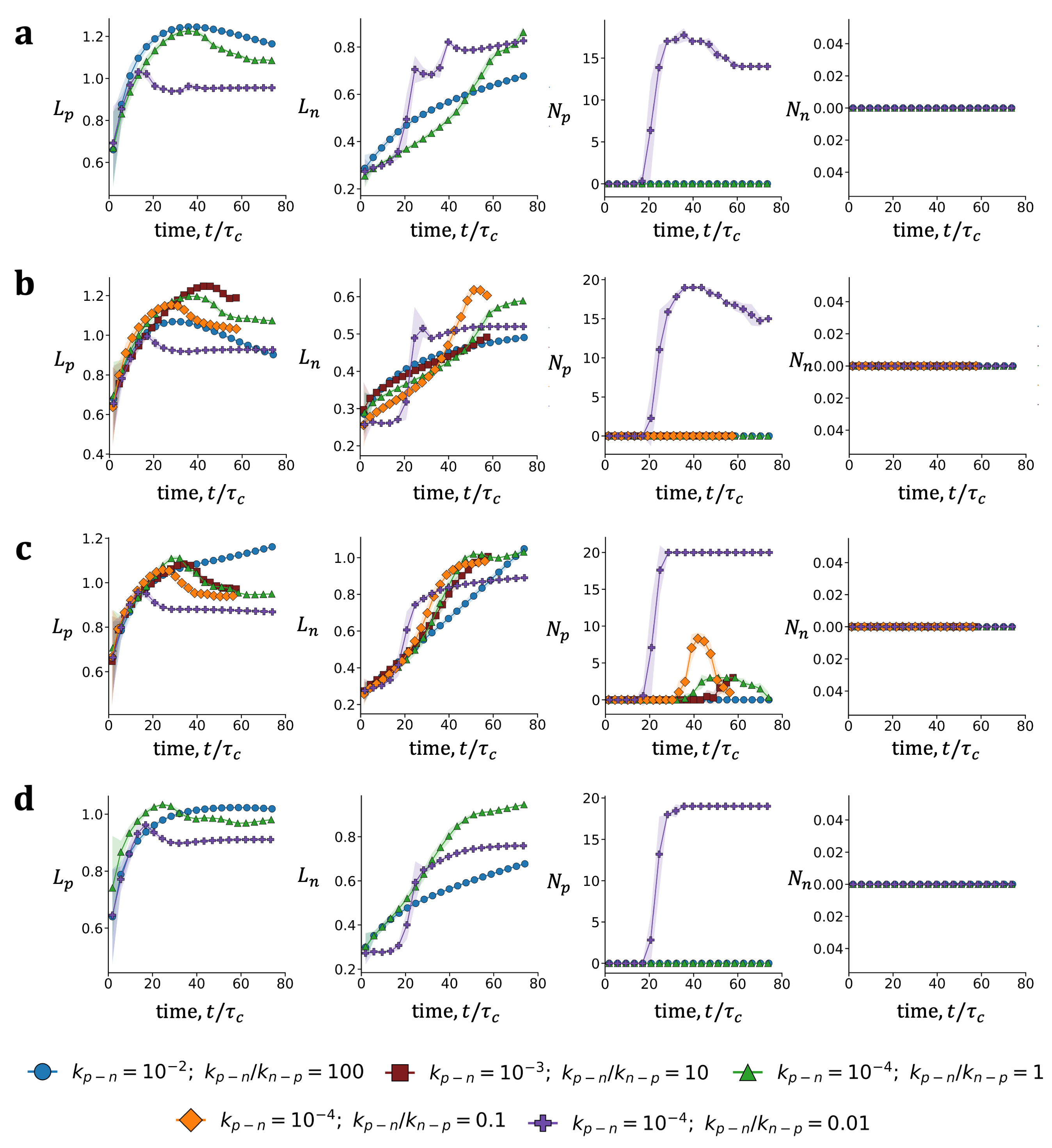}
\caption{\textbf{Morphology.} In (a)--(d), the characteristic length scales $L_p$ and $L_n$ and number of domains $N_n$ and $N_n$ of the $p-$rich and $n-$rich phases, respectively are plotted over time $t/\tau_c$ (where $\tau_c = (\delta t)_{\text{max}} = 10^{-5}$).  (a) $U_b/\kb T = 5$, $U_u/\kb T = 5$; (b) $U_b/\kb T = 5$, $U_u/\kb T = 10$; (c) $U_b/\kb T = 10 $, $U_u/\kb T = 5$; (d) $U_b/\kb T = 10$, $U_u/\kb T = 10$. These results correspond to $\alpha = 10$, $\chi_{pn} = 5$, and $\chi_{ps} = \chi_{ns} =  3$.  In (b) and (c), $N_p$ and $N_n$ represent number of phase-separating strips that reach the volume fraction 0.6 within the time frame the simulations are performed.  }
\label{fig:supp_morph_2}
\end{figure}

\begin{figure}
\centering
\includegraphics[width=0.6\linewidth]{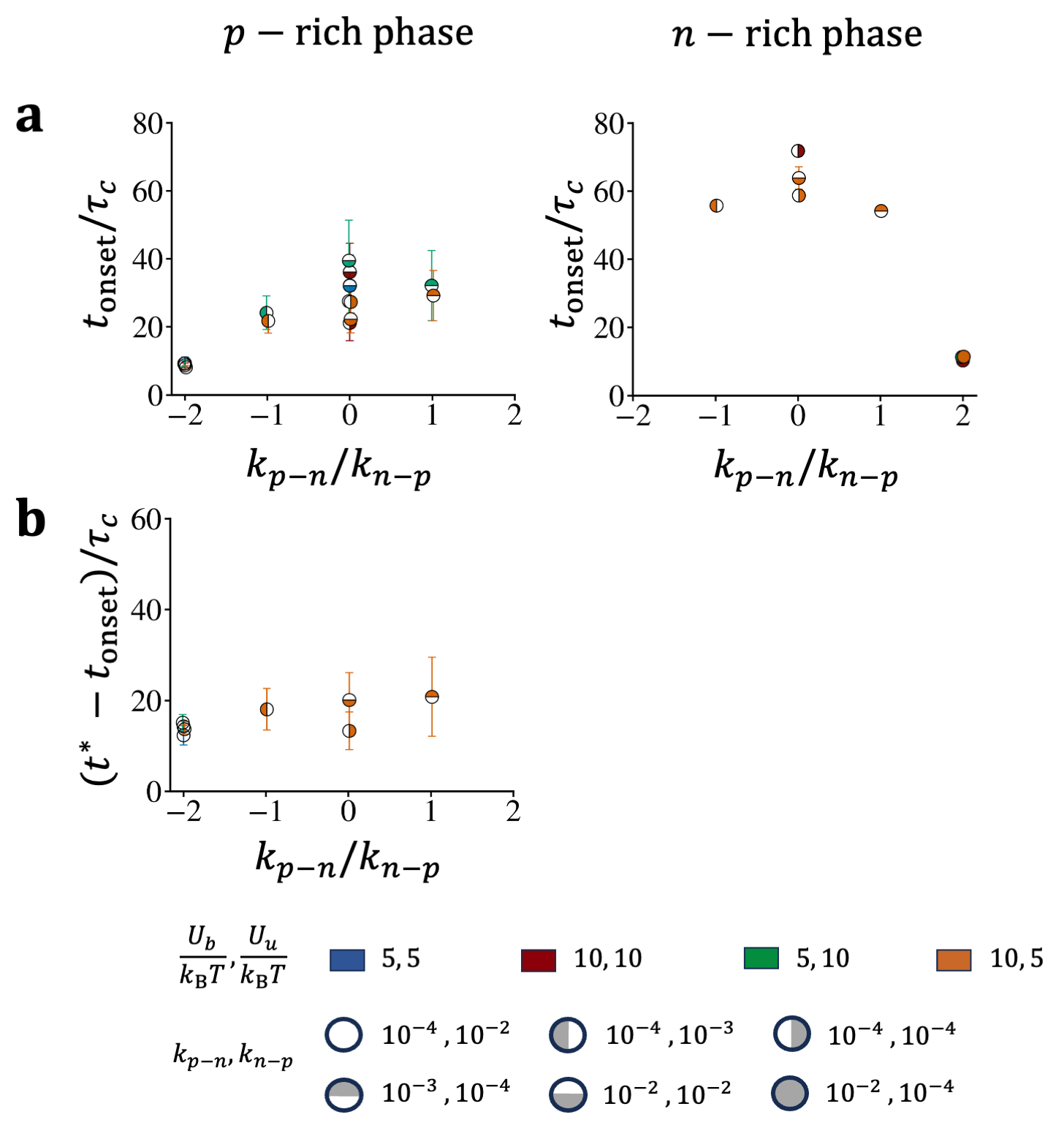}
\caption{\textbf{Onset and duration of phase separation.} The onset of phase separation $t_{\text{onset}}$ (a) and the duration of phase separation $(t^* - t_{\text{onset}})$ (b) scaled by $\tau_c$ (where $\tau_c = (\delta t)_{\text{max}} = 10^{-5}$)  are shown with $k_{p-n}/k_{n-p}$. The colors represent different combinations of $U_b$ and $U_u$, and the open, partially filled and filled circles correspond to various cases of $k_{p-n}$ and $k_{n-p}$. The figures in the left and right panels in (a) and (b) correspond to the $p-$rich and $n-$rich phases, respectively. These results correspond to $\alpha = 10$, $\chi_{pn} = 5$, and $\chi_{ps} = \chi_{ns} =  3$. Here $t_{\text{onset}}$ is obtained be averaging the time when volume fraction reaches 0.4 and $t^*$ is similarly obtained when volume fraction reaches 0.6 in the phase separating strips. }
\label{fig:time_morph_2}
\end{figure}

%\bibsplit[10]
\bibliographystyle{unsrt}
\bibliography{References}

\end{document}